\documentclass[superscriptaddress,a4paper,tightenlines,nofootinbib,floatfix]{article}
\pdfoutput=1

\usepackage{amsmath,amssymb,jheppub}
\usepackage{graphicx}
\usepackage{dcolumn}
\usepackage{bm}
\usepackage{enumitem}
\usepackage[mathlines]{lineno}
\usepackage{xcolor}
\usepackage{multirow}
\usepackage{url}
\usepackage{float,slashed}
\usepackage{soul}
\usepackage{makecell}

\usepackage[normalem]{ulem}

\def\issue(#1,#2,#3){{\bf #1}, #2 (#3)}

\catcode`\@=11
\def\lsim{\mathrel{\mathpalette\@versim<}}
\def\gsim{\mathrel{\mathpalette\@versim>}}
\def\@versim#1#2{\vcenter{\offinterlineskip
\ialign{$\m@th#1\hfil##\hfil$\crcr#2\crcr\sim\crcr } }}
\catcode`\@=12

\catcode`@=12

\newcommand{\met}{$\cancel E_T$}
\newcommand{\newc}{\newcommand}
\newc{\wt}{\widetilde}
\newc{\ra}{\rightarrow}
\def\beq {\begin{equation}}
\def\eeq {\end{equation}}
\def\bi {\begin{itemize}}
\def\ei {\end{itemize}}
\def\bea {\begin{eqnarray}}
\def\eea {\end{eqnarray}}

\def \met{\rm E{\!\!\!/}_T}

\newcommand{\br}{\begin{eqnarray}}
\newcommand{\er}{\end{eqnarray}}
\newcommand{\be}{\begin{equation}}
\newcommand{\ee}{\end{equation}}

\newcommand{\ch}{\widetilde \chi^\pm}

\def \charginopm{{\wt\chi}^{\pm}}

\def \ch2p {{\wt\chi_2^+}}

\def \ch2m {{\wt\chi_2^-}}

\def \chipm{{\wt\chi_i}^{\pm}}

\def \chonepm{{\wt\chi_1}^{\pm}}
\def \chonemp{{\wt\chi_1}^{\mp}}
\def \mchonepm{m_{\chonepm}}

\newc{\dmchi}{\Delta m_{\wt\chi}}

\def \chtwopm{{\wt\chi_2}^{\pm}}

\def \lspi{\wt\chi_i^0}

\def \lspj{\wt\chi_j^0}

\def \lspone{\wt\chi_1^0}

\def \lsptwo{\wt\chi_2^0}

\def \lspthree{\wt\chi_3^0}

\def \lspfour{\wt\chi_4^0}

\def \lspfive{\wt\chi_5^0}

\def \slep{\wt{l}}

\def \stauone{\wt\tau_1}

\usepackage{array}
\newcolumntype{L}[1]{>{\raggedright\let\newline\\\arraybackslash\hspace{0pt}}m{#1}}
\newcolumntype{C}[1]{>{\centering\let\newline\\\arraybackslash\hspace{0pt}}m{#1}}
\newcolumntype{R}[1]{>{\raggedleft\let\newline\\\arraybackslash\hspace{0pt}}m{#1}}

\def\issue(#1,#2,#3){{\bf #1}, #2 (#3)}

\hypersetup{
   colorlinks=true,       
   linkcolor=blue,        
   citecolor=red,         
   filecolor=magenta,      
   }
\title{Long live The NMSSM!}

\author[a]{Amit Adhikary}
\author[b]{Rahool Kumar Barman}
\author[c]{Biplob Bhattacherjee}
\author[c]{Amandip De}
\author[c]{Rohini M. Godbole}
\author[d]{Suchita Kulkarni}
\affiliation[a]{Institute of Theoretical Physics, Faculty of Physics, University of Warsaw, Pasteura 5, PL 02-093, Warsaw, Poland}
\affiliation[b]{Department of Physical Sciences, Oklahoma State University, Stillwater, Oklahoma 74078, USA}
\affiliation[c]{Centre for High Energy Physics, Indian Institute of Science, Bengaluru - 560012, India}
\affiliation[d]{Institute of Physics, NAWI Graz, University of Graz, Universit\"atsplatz 5, A-8010 Graz, Austria}
\emailAdd{amit.adhikary@fuw.edu.pl}
\emailAdd{rahool.barman@okstate.edu}
\emailAdd{biplob@iisc.ac.in}
\emailAdd{rohini@iisc.ac.in} 
\emailAdd{amandipde@iisc.ac.in}
\emailAdd{suchita.kulkarni@uni-graz.at}

\abstract{We analyze the scenario within the Next to Minimal Supersymmetric Standard Model (NMSSM), where the lightest supersymmetric particle (LSP) is singlino-like neutralino. By systematically considering various possible admixtures in the electroweakino sector, we classify regions of parameter space where the next to lightest supersymmetric particle (NLSP) is a long-lived electroweakino while remaining consistent with constraints from flavor physics, dark matter direct detection, and collider data. We identify viable cascade decay modes featuring the long-lived NLSP for directly produced chargino-neutralino pairs, thus, leading to displaced vertex signatures at the high luminosity LHC (HL-LHC). We construct track based analysis in order to uncover such scenarios at the HL-LHC and analyze their discovery potential. We show that through such focused searches for the long-lived particles at the HL-LHC, one can probe regions of the electroweakino parameter space that are otherwise challenging.} 

\begin{document}

\maketitle
\section{Introduction}
\label{sec:intro}

Observations of the existence and measurements of dark matter (DM)~\cite{Planck:2018vyg,ParticleDataGroup:2018ovx}, of matter-antimatter asymmetry and non-zero neutrino masses,~\cite{ParticleDataGroup:2018ovx} as well as theoretical considerations such as the hierarchy problem~\cite{Gildener:1976ih,Susskind:1978ms,tHooft:1979rat} all point to the existence of new physics beyond that in the Standard Model(SM). Among the extensions of the SM, those involving supersymmetry (SUSY) still remain one of the most appealing because they address multiple shortcomings of the SM at once~\cite{WESS197439, NILLES19841, HABER198575, doi:10.1142/4001, baer_tata_2006}. Depending on the exact realization, SUSY can present numerous dark matter (DM) candidates such as the lightest neutralino, sneutrino, or gravitino~\cite{doi:10.1142/4001,Bertone:2004pz,Belanger:2009br,KumarBarman:2020ylm}. The SUSY DM candidate, viz. the Lightest Supersymmetric Particle (LSP), is stabilized by means of an external symmetry, such as R-parity, see, for example, \cite{doi:10.1142/4001, baer_tata_2006}.

The interactions of the DM with the particles in the SM or those within the dark sector affect its exact evolution and hence subsequently, the prediction for the amount remaining today, dubbed as relic density. This has been now accurately measured to be $\Omega h^2 = 0.120 \pm 0.001$~\cite{Planck:2018vyg} where $h$ is Hubble constant in units of $100 {\rm ~km ~s^{-1} Mpc^{-1}}$. In case of the neutralino LSP, the relic density is often generated by means of the popular thermal freeze-out process~\cite{PhysRevLett.39.165,Moroi:1999zb,Garrett:2010hd}, while for gravitino, the suppressed couplings with the SM necessitate a non-thermal relic density generation mechanism~\cite{PhysRevLett.48.1303,ELLIS1985175}. Apart from the particle physics aspect, the relic density also depends on the details of the early Universe evolution. For example, late time entropy production can substantially dilute the relic density while keeping the particle physics details unchanged. In the absence of precise knowledge of DM interactions and the evolution of the early Universe, it is thus important to consider both over-abundant and under-abundant (viz. with predicted relic density has a value above (below) the measured value) regions of SUSY DM parameter space~\cite{Gelmini:2006pw, Baer:2014eja, Aparicio:2015sda, Aparicio:2016qqb, Roszkowski:2017dou, Barman:2017swy, KumarBarman:2020ylm}.

Dark matter can be searched for in several experiments. Due to model independent search strategies, the results are applicable to SUSY and a variety of other beyond the SM scenarios. The primary detection strategies are via detection of missing energy at the LHC, via scattering off nuclei at underground direct detection experiments, or via detection of decay or annihilation products through cosmic rays in the Universe today at indirect detection experiments. Among these, the direct detection experiments already rule out left-handed sneutrino DM arising in the minimal supersymmetric Standard Model~\cite{Falk:1994es}. Out of the thermal candidates, this leaves the lightest neutralino -- a linear combination of the bino, wino, and higgsino -- as a viable DM candidate, whose compatibility with the experimental searches needs to be checked in detail.

In the MSSM, the lightest neutralino is a part of the system of electroweakinos, which consists of four neutralinos and two charginos. The electroweakino sector, and in particular, light neutralinos, have been a topic of intense phenomenological and experimental investigations in the past decade.  Some of the latest LHC results for electroweakino searches are summarized below. A CMS search for electroweakinos through chargino-neutralino production ($\chonepm \lsptwo$) with on-shell decays to $Wh$ final state rules out wino-like chargino masses up to 700 GeV, for bino-like LSP mass $M_{\lspone} < 350 $ GeV~\cite{CMS:2021few}. This search was performed at the center-of-mass energy of 13 TeV with an integrated luminosity of 137 ${\rm fb^{-1}}$. Another search from the ATLAS collaboration considers pair production of neutralinos at 13 TeV with 139 fb$^{-1}$ integrated luminosity in fully hadronic final states mediated by $WW, WZ$ or $Zh$~\cite{PhysRevD.104.112010}. This search excludes wino (higgsino) mass up to 1060 (900) GeV for bino-like LSP up to 400 (200) GeV. These searches imply a relatively heavier electroweakino sector. It should, however, be noted that these results assume a simplified model framework with 100\% branching ratios, which should be reinterpreted in the context of specific SUSY models, e.g., pMSSM or NMSSM. As a result, lighter electroweakinos can still be allowed despite the stringent LHC limits, and the exact limits are model dependent.

Some generic conclusions about the MSSM neutralino dark matter in light of recent collider and astrophysical constraints are available now. For example, the neutralino masses in phenomenological MSSM (pMSSM) have a lower limit of ${\rm M_{\lspone} > 34 ~ GeV}$ in order to avoid over-abundant relic density~\cite{Calibbi:2013poa,Belanger:2013pna,Barman:2017swy,Cahill-Rowley:2014twa,Cao:2015efs}. In the general-MSSM scenario, higgsinos are favored to have mass $\sim$ 1 TeV, to obtain the correct DM abundance for a single component thermal DM~\cite{Arkani-Hamed:2006wnf,Baer:2016ucr,Chakraborti:2017dpu}. Within the MSSM, relic density compliant regions require either heavy DM or rely on a co-annihilation mechanism, which demands a small mass splitting between DM and its co-annihilating partner. Such small mass gaps can lead to long-lived particles (LLP), which can then be investigated, for example, by looking for displaced vertices or heavy stable charged particles. It is worth noting that relaxing the DM relic density requirement does not necessarily lead to additional LLP parameter space within the MSSM. This is because the only way to obtain LLPs is through small mass splitting, as the SUSY couplings are related to those of SM and hence can not be suppressed.

Although the MSSM can successfully provide a DM candidate, a drawback of this most commonly used SUSY realization is ``$\mu$-problem" which arises as an artifact of the common mass term for two Higgs doublets. This introduces a fine-tuning, which requires an electroweak scale $\mu$ parameter rather than the expected Planck scale~\cite{Kim:1983dt}. An alternative can be considered as a singlet extension of the MSSM, the next-to-minimal supersymmetric standard model (NMSSM)~\cite{NILLES1983346,PhysRevD.39.844,Ellwanger:1993xa} with a singlet Higgs field in addition to the two Higgs doublets of MSSM. For this additional scalar, the effective $\mu$ term can be generated dynamically, alleviating the fine-tunning of $\mu$. The fermionic component of the singlet superfield provides an additional neutralino without violating the existing constraints. In such cases, the LSP can be pure singlino dominated or a mixture of higgsino-singlino. Such LSP can be lighter than the corresponding MSSM counterpart~\cite{Abel:1992ts,Kozaczuk:2013spa,Cao:2013mqa,Han:2014nba,Ellwanger:2014dfa}. 

The phenomenology of such extended sectors can open up interesting new avenues for DM phenomenology as well as experimental searches. In this work, we revisit the neutralino sector of the NMSSM, focusing on the LSP with a significant singlino fraction~\cite{Cao:2021ljw,Zhou:2021pit,Barman:2020vzm,Guchait:2020wqn}. Such singlino has suppressed couplings with the rest of the SUSY spectrum and thus can lead to a long-lived NLSP neutralino. We investigate this possibility and suggest displaced vertex search relying on tracks originating through NLSP decays. It should be noted that the region of the LLP parameter space in the NMSSM has two distinct features. First, the LLPs are a result of suppressed couplings and not small mass differences, and second, a large part of the LLP parameter space corresponds to over-abundant relic density.

The LLPs themselves are intriguing since they lead to characteristic signatures at the colliders. The charge and color neutral LLPs travel a macroscopic distance before decaying into SM particles at a secondary vertex, resulting in a displaced vertex signature. The LLPs can be realized either with scenarios involving suppressed couplings or small mass splittings. Depending on the LLP lifetime, its decay may take place either in the tracker, or in calorimeters and muon system, or even outside the detector. The pivotal advantage is having an almost negligible background, thanks to the existence of displaced vertices. A variety of theory scenarios, including SUSY, little Higgs~\cite{Cai:2008au}, twin Higgs~\cite{Chacko:2005pe}, dark sector models~\cite{Baumgart:2009tn,Kaplan:2009ag,Dienes:2011ja,Dienes:2012yz,Alimena:2019zri} etc, predict LLPs. In SUSY, LLPs are usually featured in R-parity violating models~\cite{Barbier:2004ez}. Besides, in many R-parity conserving (RPC) scenarios gauge-mediated SUSY (GMSB)~\cite{Dimopoulos:1996vz,Giudice:1998bp}, anomaly mediated SUSY (AMSB)~\cite{Feng:1999fu}, particles with long lifetime can appear.

The long-lived NLSP neutralino within the NMSSM is thus an exciting prospect, and a potential discovery could lead to a renewed understanding of the behavior of dark matter in the early Universe. We, therefore, present a detailed search strategy for such parameter space in this work. The rest of the paper is organized as follows. In Sec. \ref{sec:NMSSM_framework} we briefly review the NMSSM framework and motivation for the relevant parameters to single out the region of interest. Sec. \ref{sec:parameter_space} describes the pertinent range of parameters for numerical scan, along with the current phenomenological constraints. The characteristic features of the parameters to achieve long-lived neutralinos are discussed in Sec. \ref{sec:allowed_parameter_space}. In Sec. \ref{sec:Result} we present a signal-to-background study via searches of displaced vertices from decays of long-lived neutralinos and explore the reach of such searches for direct production of electroweakinos at the HL-LHC. Finally, we conclude in Sec.~\ref{sec:conclusion}.

\section{The NMSSM framework}
\label{sec:NMSSM_framework}
\subsection{Higgs sector}

In this section, we discuss the Higgs and electroweakino sectors in the NMSSM. The NMSSM Higgs sector consists of a singlet superfield $\hat{S}$ and two doublet Higgs superfields, $\hat{H}_{u}$ and $\hat{H}_{d}$. The dimensionful couplings of $\hat{S}$ can be forbidden through a discrete $\mathcal{Z}_{3}$ symmetry leading to scale invariant NMSSM superpotential~\cite{Ellwanger_2010}
\begin{equation}
   W_{NMSSM} = W_{MSSM}(\mu =0) + \lambda \hat{S}\hat{H}_u\cdot\hat{H}_d + \frac{1}{3}\kappa \hat{S}^3.
   \label{eq:NMSSM_superpotential}
\end{equation}
Here, $W_{MSSM}(\mu =0)$ is the MSSM superpotential without the $\mu$-term, while $\lambda$ and $\kappa$ are dimensionless couplings. The $\lambda \hat{S}\hat{H}_u.\hat{H}_d$ term generates an effective MSSM-like $\mu$-term when $\hat{S}$ develops a vacuum expectation value~(vev) $v_s$, $\mu = \lambda v_s$. Thus, the $\mu$-term in NMSSM is generated `dynamically', providing a solution to the MSSM $\mu$-problem~\cite{Kim:1983dt} when $v_s$ is at the electroweak scale~\cite{Ellwanger:2009dp}. The soft SUSY breaking terms containing the singlet and doublet Higgs fields have the form
\begin{equation}
\begin{split}
   V_{soft} = & ~m^2_{H_u}|H_u|^2 + m^2_{H_d}|H_d|^2 + m^2_{S}|S|^2 \\ 
   &+ \left( \lambda A_\lambda S H_u\cdot H_d  + \frac{1}{3} \kappa A_k S^3 + \text{h.c.} \right),
\end{split}
   \label{Eqn:V_soft_term}
\end{equation}
where, $m_{H_u}$, $m_{H_d}$, $m_{S}$ are the soft breaking Higgs masses, and $A_{\lambda}$, $A_{\kappa}$ are the trilinear couplings. The Higgs scalar potential $V$ is also augmented by F- and D-terms, 
\begin{equation}
\begin{split}
    &V_F = |\lambda H_u\cdot H_d + \kappa S^2|^2 + \lambda^2 |S|^2 \left( H^\dagger_u H_u  + H^\dagger_d H_d\right), \\ 
    &V_D = \frac{g_1^2 + g_2^2 }{8} \left(  H^\dagger_u H_u  -  H^\dagger_d H_d\right) + \frac{g_2^2}{2} |H^\dagger_d H_u|^2,
\end{split}    
\label{Eqn:D_term}
\end{equation}
respectively. In Eq.~\eqref{Eqn:D_term} $g_{1}$ and $g_{2}$ are the SM $U(1)_{Y}$ and $SU(2)_{L}$ gauge couplings, respectively. The physical Higgs states $\{H_{u}^{0}, H_{d}^{0}, S\}$ can be obtained by expanding the Higgs scalar potential in Eq.~\eqref{Eqn:V_soft_term}~-~\eqref{Eqn:D_term}, $V_{soft}+V_{D}+V_{F}$, around real neutral $vevs$ $v_{u},~v_{d}$ and $v_{s}$, and following the notation of~\cite{Baum:2017gbj}, are given by,
\begin{equation}
\begin{split}
    &H^0_u = \frac{v_u + H_u^R + iH_u^I}{\sqrt{2}},\\
    &H^0_d = \frac{v_d + H_d^R + iH_d^I}{\sqrt{2}},~S = \frac{v_s + H^S + iA^S}{\sqrt{2}}.
\end{split}
\end{equation}
Here, $\{H_u^R,~H_d^R,~H^S\}$ are the real components while $\{H_u^I,~H_d^I,~A^S\}$ are the imaginary components. The three real components lead to three neutral CP-even Higgs bosons. One neutral pseudoscalar Higgs boson stems from the imaginary components $\{H_u^I,~H_d^I\}$, while $\{A^S\}$ leads to another neutral pseudoscalar Higgs boson. The mass matrix elements of CP-even Higgs can be computed in a rotated CP-even Higgs interaction basis $\{H^{SM},H^{NSM},H^{S}\}$ where $H^{SM}$, $H^{NSM}$ and $H^{S}$ corresponds to SM-like, MSSM-like heavy Higgs and singlet scalar Higgs eigenstates, respectively. The elements of the $3\times 3$ symmetric mass-squared matrix $M_{S}^{2}$ in this basis are given by~\cite{Baum:2017gbj,Ellwanger:2009dp,PhysRevD.93.035013},
\begin{equation}
\begin{split}
M^2_{S,11} &= \left(m^2_Z - \frac{1}{2}\lambda^2\right)\sin2\beta^2 + \frac{\mu}{\sin\beta \cos\beta}\left(A_\lambda + \frac{\kappa\mu}{\lambda}\right),\\
M^2_{S,22} &= m_Z^2\cos2\beta^2 + \frac{1}{2}\lambda^2v^2\sin2\beta^2,\\
M^2_{S,33} &= \frac{1}{4}\lambda^2v^2\sin2\beta\left(\frac{A_\lambda}{\mu}\right) + \frac{\kappa\mu}{\lambda}\left(A_\kappa + \frac{4\kappa\mu}{\lambda}\right),\\
M^2_{S,12} &= \left(\frac{1}{2}\lambda^2v^2 - m^2_Z\right)\sin2\beta\cos2\beta,\\
M^2_{S,13} &= -\frac{1}{\sqrt{2}}\lambda v\cos2\beta\left(\frac{2\kappa\mu}{\lambda} + A_\lambda\right),\\
M^2_{S,23} &= \sqrt{2}\lambda v \mu\left(1 - \frac{A_\lambda}{2\mu}\sin2\beta - \frac{\kappa}{\lambda}\sin2\beta\right),\\
\end{split}
\end{equation}
Here, $\beta = \tan^{-1} \frac{v_{u}}{v_{d}}$ with $\sqrt{v_{u}^{2} + v_{d}^{2}} = v \simeq 246{\rm ~GeV}$ and $m_Z$ represents the Z boson mass. The CP-even Higgs mass eigenstates $H_i~(i=1,2,3)$ can be obtained by diagonalizing $M_{S}^{2}$ through an orthogonal rotation matrix $V$; $H_i = \sum\limits_{j=1}^3 V_{i,j}\phi_{jR}$, where $\phi_{R} = \{H^{SM},H^{NSM},H^{S}\}$, and $m_{H_{1}} < m_{H_{2}} < m_{H_{3}}$. In the present study, we require $H_{1}$ to be consistent with the properties of the observed 125~GeV Higgs boson. Introducing  
\begin{equation}
    M^2_A = \frac{\mu}{\sin\beta\cos\beta}\left(A_\lambda + \frac{\kappa\mu}{\lambda}\right)
\end{equation}
the elements of the $2\times 2$ symmetric mass-squared matrix $M_{P}^{2}$ after dropping the Goldstone modes in the pseudoscalar Higgs interaction basis $\phi_{I} = \{A^{NSM}, A^S\}$ can be written as,
\begin{equation}
\begin{split}
M^2_{P,11} &= M^2_A,\\
M^2_{P,22} &= \frac{1}{2}\lambda^2 v^2 \sin2\beta \left(\frac{M^2_A}{4\mu^2}\sin2\beta + \frac{3\kappa}{2\lambda}\right) - \frac{3\kappa A_\kappa \mu}{\lambda},\\
M^2_{P,12} &= - \frac{1}{\sqrt{2}} \lambda v \left( \frac{3\kappa \mu}{\lambda} - \frac{M^2_A}{2\mu}\sin2\beta\right) 
\label{eqn:pseudoscalar_mass}
\end{split}
\end{equation}
Here again, following a similar recipe, the CP-odd Higgs mass eigenstates $\{A_{m} = A_{1},A_{2}\}$ ($m_{A_{1}} < m_{A_{2}}$) can be written as $A_{m} = \sum\limits_{n=1}^2 P_{m,n}\phi_{nI}$, where $m=1,2$ and $P_{m,n}$ is an orthogonal rotation matrix. 

In addition to the three CP-even and the two CP-odd neutral Higgs states, the NMSSM framework also predicts a pair of charged Higgs bosons $H^{\pm}$. At tree level, their masses are given by, 
\begin{equation}
    M^2_{H^\pm} = M^2_A + m_W^2 - \frac{1}{2}\lambda^2v^2
\end{equation}
where $m_W$ is the mass of W boson. Overall, the tree level Higgs sector of NMSSM can be parametrized by the following 6 parameters:
\begin{equation}
    \lambda,\kappa,A_\lambda,A_\kappa,\tan\beta,\mu
    \label{higgs_params}
\end{equation}

\subsection{Electroweakino sector}
\label{sec:electroweakino_sector}

The NMSSM electroweakino sector consists of bino $\tilde{B}^0$, neutral wino $\tilde{W}_3^0$, higgsinos $\tilde{H}^0_d$, $\tilde{H}^0_u$, and singlino $\tilde{S}$, leading to 5 neutralino and 2 chargino mass eigenstates. In the $\{\tilde{B},\tilde{W}_3^0,\tilde{H}^0_d, \tilde{H}^0_u, \tilde{S}\}$ basis, the symmetric 5$\times$5 neutralino mass matrix $M_{\tilde{N}}$ can be written as,

\begin{equation}
\resizebox{.9\hsize}{!}{$
M_{\tilde{N}} = 
\begin{pmatrix}
M_1 & 0 & -m_Z \sin\theta_W \cos\beta & m_Z \sin\theta_W \sin\beta & 0 \\
0 & M_2 & m_Z \cos\theta_W \cos\beta & -m_Z \cos\theta_W \sin\beta & 0 \\
-m_Z \sin\theta_W \cos\beta & m_Z \cos\theta_W \cos\beta & 0 & -\mu  & -\lambda v \sin\beta\\
m_Z \sin\theta_W \sin\beta & -m_Z \cos\theta_W \sin\beta & -\mu & 0  & -\lambda v \cos\beta\\
0 & 0 & -\lambda v \sin\beta & -\lambda v \cos\beta & 2\kappa v_s 
\end{pmatrix}
\label{mass_mat_nu}
$}
\end{equation}

Here, $M_{1}$ is the bino mass parameter, $M_{2}$ is the wino mass parameter, and $\theta_W$ is the Weinberg angle. Diagonalizing $M_{\tilde{N}}$ through an $5\times 5$ orthogonal rotation matrix $\hat{N}$ leads to the neutralino mass eigenstates $\tilde{\chi}_{i}^{0}$,
\begin{equation}
\lspi = \hat{N}_{i1}\tilde{B}^0 +  \hat{N}_{i2}\tilde{W}_3^0 + \hat{N}_{i3}\tilde{H}_d^0 + \hat{N}_{i4}\tilde{H}_u^0 + \hat{N}_{i5}\tilde{S}   
\end{equation}
Similarly, the charged winos and higgsinos mix to generate the two charginos $\tilde{\chi}_{i}^{\pm}~(i=1,2)$. The input parameters that regulate the electroweakino sector at the tree level are as follows: 
\begin{equation}
    M_1, M_2, \mu, \tan\beta, \lambda, \kappa
\label{ew_params}     
\end{equation}

The lightest neutralino $\lspone$ naturally provides a DM candidate in R-parity conserving NMSSM. \textit{A priori}, the LSP $\lspone$ can be pure gaugino, higgsino, singlino, or an admixture of these states. Such an LSP can lead to correct DM relic density either if it is purely higgsino or wino-like with masses up to $2.8~\mathrm{TeV}$~\cite{Fan:2013faa, Bramante:2015una} or if it is bino or singlino like which can annihilate through co-annihilation or resonant annihilation through Higgs or $Z$ boson. Such co-annihilation conditions can only be realized for $m_{\lspone} \sim m_{H/Z}/2$, subject to non-negligible $\lspone\lspone Z/H$ couplings. In this work, however, we do not impose any relic density requirements and consider both under-abundant and over-abundant regions of parameter space. For these scenarios, the relic density can be fulfilled either by requiring additional DM candidates or by requiring non-standard evolution of the Universe, as argued in Sec~\ref{sec:intro}.


\section{Parameter space scan and constraints}
\label{sec:parameter_space}

Our primary focus is the region in the parameter space of the RPC NMSSM, featuring a long-lived neutralino while being consistent with the current collider and direct/indirect detection constraints. To this end, we consider a dominantly singlino-like LSP $\lspone$ and bino-like NLSP $\lsptwo$. Since there are no tree level couplings between the bino and singlino, the bino-like NLSP $\lsptwo$ decays to the singlino-like LSP $\lspone$ only through their mutual higgsino admixtures. This leads to a suppressed coupling between LSP and NLSP states. An additional phase space suppression can be achieved if the mass difference between the two states is smaller than the $Z$ mass. In such scenarios, the bino-like $\lsptwo$ can be LLP. The heavier neutralinos $\lspthree, \lspfour, \lspfive$, and charginos $\chonepm, \chtwopm$ can be either higgsino-like, wino-like, or admixtures of both and decay promptly. In this analysis, we consider a moderately mixed scenario with $\mu < M_{2}$ which implies a relatively large higgsino admixture in $\lspthree, \lspfour$, and $\chonepm$. 

Our choice for $\mu < M_{2}$ is motivated by three factors. First, LHC constraints for higgsinos are weaker compared to winos~\cite{CMS:2021few,PhysRevD.104.112010}. Second, higgsinos have tree-level couplings with both singlino and bino, while no such interactions exist for wino-bino or wino-singlino. Therefore, winos can decay into bino or singlino only by virtue of its mixing with higgsinos. Third, both bino-like $\lsptwo$ and singlino-like $\lspone$ are required to have non-zero higgsino admixtures in order to generate a tractable decay width for $\lsptwo$ such that they can be probed at the LHC through track-based LLP searches. Concretely, we choose $500~\mathrm{GeV} \lesssim \mu \lesssim 1000~\mathrm{GeV}, M_{2}\geq 2~\mathrm{TeV}$ such that $\lspthree, \lspfour$, and $\chonepm$ have a dominant higgsino admixture with appreciable production rates at HL-LHC, compatible with existing LHC constraints from direct electroweakino searches, discussed in Sec.~\ref{sec:constraints}.

In the NMSSM superpotential as given in Eq.~(\ref{eq:NMSSM_superpotential}), we observe that interactions between the singlet superfield $\hat{S}$ and the MSSM Higgs superfields $\hat{H}_{u},\hat{H}_{d}$ is controlled by $\lambda$. In the limit, $\lambda \to 0$~(for a fixed $\mu = \lambda v_{S}$), the singlet-like scalar, singlet-like pseudoscalar, and the singlino can no longer interact with the MSSM sector. This consideration leads to the possibility of a pure singlino-like neutralino LSP with a tree level mass $\sim 2\kappa v_{S}$. In this case, the NLSPs would be composed of bino/wino/higgsinos, similar to that in MSSM. Furthermore, in the $\lambda \to 0$ limit, the LSP has no interaction with NLSPs, however, keeping a finite but small $\lambda$ leads to suppressed interactions between singlino LSP and MSSM-like neutralino NLSPs. This suppression leads to long-lived NLSPs, which is the focus of this work. In particular, we consider bino-like $\lsptwo$. In the limit, $\mu \gg 2\kappa v_{S}$, the mass of the singlino-like neutralino $m_{\lspone}$ can be approximated as:
\be
m_{\lspone}\sim 2\kappa v_{S} \simeq 2 \frac{\kappa}{\lambda} \mu.
\label{eq:singlino_mass}
\ee
We therefore observe that a singlino-like LSP with a typical mass of $\mathcal{O}(100)~\mathrm{GeV}$ and $\mu \sim \mathcal{O}(500)~\mathrm{GeV}$ leads to $\kappa/\lambda \sim \mathcal{O}(0.1)$. In order to maintain a similar mass hierarchy between the higgsino-like neutralinos and the singlino-like $\lspone$, we restrict ourselves to $\kappa/\lambda \leq 0.15$ with $10^{-5} \leq \lambda \leq 10^{-1}$. Correspondingly, for the sake of simplicity, we restrict ourselves to the parameter space where singlino-like LSP mass is $\mathcal{O}(100)~\mathrm{GeV}$. We are thus left with bino mass parameter, $M_{1}$ the only remaining parameter in the electroweakino mass spectrum which is not fixed. Since, we are interested in a bino-like $\lsptwo$, it must fall between the singlino- and higgsino-like neutralinos. Correspondingly, we vary $M_{1}$ over the range $150~\mathrm{GeV} \leq M_{1} \leq 550~\mathrm{GeV}$. The other input parameters that are relevant to the present study are: $A_{\lambda}$, $A_{\kappa}$, the gluino mass parameter $M_{3}$, squark mass parameters $M_{U_R,D_{R}}^{i}$, $M_{{Q}_{L}}^{i}$~(i=1,2,3), the tri-linear couplings $A_{t}$, $A_{b}$, $A_{\tau}$, the slepton mass parameters $M_{E}^{i}$, $M_{L}^{i}$. We set $A_{b}$, $A_{\tau}$, squark and slepton mass parameters to $2~\mathrm{TeV}$. In order to maximize the one-loop top/stop contributions to the lightest CP-even Higgs mass the tri-linear soft coupling  $A_t$ is varied over a wide range [-10,10] TeV. To respect the charge and color breaking minima~\cite{Kusenko:1996jn,Chowdhury:2013dka}, we exploit the maximum mixing scenario (c.f. Ref.~\cite{Carena:2000dp}) and require that the ratio of stop mixing parameter $|X_{t}|$ ($= A_{t} - \mu \cot\beta$) to average stop mass $M_{T}$ ($ M_{T}^{2}= m_{\tilde{t_{1}}}m_{\tilde{t_{2}}}$, where $m_{\tilde{t}_{1},\tilde{t}_{2}}$ are the stop masses) to satisfy $|X_t/M_T| < 2.5$~\cite{Brummer:2012ns,Chowdhury:2013dka}.

\subsection{Scan range}

We utilize the \texttt{NMSSMTools-5.5.3}~\cite{Ellwanger_2005, Ellwanger_2006} package to perform a random scan over the parameter space. The particle masses, couplings, branching ratios, and decay widths are also computed using \texttt{NMSSMTools-5.5.3}. We perform a flat random scan over $10^{8}$ points. The efficiency for obtaining allowed parameter space consistent with current collider and astrophysical data~(discussed in Sec.~\ref{sec:constraints}) is 0.001~$\%$. The scan is performed over the following range of input parameters: 

\begin{equation}
\begin{split}
&10^{-5} < \lambda < 0.1 ,~\left|\frac{\kappa}{\lambda}\right|\leq 0.15,~M_1 = (150,550)~{\rm GeV},\\
&M_2=(2000,3000)~{\rm GeV},~M_3= (3000, 10000)~{\rm GeV},\\
&\mu= (500,1000)~{\rm GeV},~\tan\beta = (1,40),\\
&A_\lambda = (-100,10000)~{\rm GeV},~A_\kappa = (-1000,100)~{\rm GeV},\\
&A_{t} = (-10000, 10000)~\mathrm{GeV}\, \\ 
\label{eq:param_scan}
\end{split}
\end{equation}

\subsection{Constraints}
\label{sec:constraints}
As discussed previously, the lightest CP-even Higgs boson $H_{1}$ plays the role of the observed SM-like Higgs boson. In this regard, $H_{1}$ is required to be consistent with the Higgs mass constraints and Higgs signal strength constraints coming from the LHC. The heavier CP-even Higgs bosons $H_{2}, H_{3}$ and the CP-odd Higgs bosons $A_{1}, A_{2}$ can be an admixture of singlet and doublet components and can be constrained by heavy Higgs searches at the LHC. The constraints from heavy Higgs searches are subject to the doublet content and get weaker with increasing singlet admixture. Furthermore, the NMSSM parameter space of our interest is also constrained by limits from LEP searches, flavor physics, direct and indirect detection experiments, and direct electroweakino searches at the LHC. We discuss various constraints below.
\begin{itemize}

\item {\bf Mass of SM-like Higgs boson:} The mass of the observed Higgs boson has been measured to be within $124.4$-$125.8$~GeV at $3\sigma$ uncertainty~\cite{ATLAS:2015yey}. Acknowledging the theoretical uncertainties in Higgs mass computation~\cite{Allanach:2004rh,Heinemeyer:2007aq,Borowka:2015ura}, and adopting a conservative approach, we allow $m_{H_{1}}$ to lie within the range $122~\mathrm{GeV} \leq m_{H_{1}} \leq 128~\mathrm{GeV}$.
	
\item {\bf Limits from LEP}: We impose a lower limit on the chargino mass $M_{\chonepm} > 103.5$~GeV which implies a lower bound of $\mu,M_{2} \gtrsim 100~\mathrm{GeV}$ \cite{LEP:CharginoMass}. Searches at LEP have also derived an upper limit of $0.1~$pb on the production cross-section of $e^{+}e^{-} \to (\lsptwo \to q\tilde{q} \lspone)\lspone$ for $|m_{\lsptwo} - m_{\lspone}| > 5~\mathrm{GeV}$~\cite{OPAL:2003wxm}. We also require $\Gamma_{Z_{\mathrm{inv}}} < 2~\mathrm{MeV}$~\cite{ALEPH:2005ab}, where $\Gamma_{Z_{\mathrm{inv}}}$ is the invisible decay width for the $Z$ boson excluding neutrinos. These constraints have been imposed using the \texttt{NMSSMTools-5.5.3} package.
	
\item {\bf Constraints from Higgs signal strength measurements:} Measurements by the ATLAS and CMS collaborations of the couplings of the 125~GeV Higgs boson with SM particles are encoded via signal strength parameters $\mu^{f}_{i}$ defined as,
	\begin{equation}
	    \mu^f_i = \frac{\sigma_i \times \text{BR}^f} {(\sigma_i)_{\text{SM}} \times (\text{BR}^f)_{\text{SM}}}.
	\end{equation}
Here, $i$ represents the various production modes of the Higgs boson $viz$ gluon fusion~($ggF$), vector boson fusion~($VBF$), associated production with vector bosons~($VH$~($V = W^{\pm}, Z$)), while $f$ denotes the various decay modes of the Higgs $viz$ $b\bar{b}, \gamma\gamma, \tau^+\tau^-, ZZ^*, W^+W^{*-}$. In the present analysis, signal strength measurements for $\{i=ggF, f = \gamma\gamma, W^+ W^-, b \bar{b}, \tau \tau, ZZ\}$, $\{i=VBF, f = \gamma\gamma, \tau \tau\}$, and $\{i=VH, f = W^+ W^-, b \bar{b}\}$~\cite{CMS:2021kom, CMS-PAS-HIG-19-017, CMS:2020dvg, CMS:2018nsn, CMS-PAS-HIG-19-010, CMS:2017dib}, have been implemented through a global $\chi^{2}$ fit assuming $2\sigma$ uncertainty following the strategy in Ref.~\cite{Barman:2016jov}.

\item {\bf Constraints from flavor physics:} The flavor physics constraints are imposed through bounds on the branching ratios of relevant rare processes $viz$ $Br(b \to s\gamma)$, $Br(B_{s} \to \mu^{+}\mu^{-})$ and $Br(B^{+} \to \tau^{+}\nu_{\tau})$, which are sensitive to SUSY contributions. For example, the loop-mediated $b\to s\gamma$ process is sensitive to modifications from loops involving charged Higgs-top, neutral Higgs-bottom, and electroweakino-squark. Contributions from the latter decouple since the squark masses have been fixed at a rather high value $\sim 2~\mathrm{TeV}$. The $B_{s} \to \mu^{+}\mu^{-}$ process is mediated through penguin and box diagrams at one loop. Both contributions are sensitive to a loop containing scalar or pseudoscalar heavy Higgs and a down quark.
The contributions from the penguin diagram are also sensitive to modifications from loops containing charged Higgs-up quark, higgsino-up quark, and gaugino-up quark, while loops from up quark-charged Higgs-neutrino, up squark-charged higgsino-sneutrino and up squark-charged wino-sneutrino can induce modifications to the box diagram contributions.
Recent measurements are: $Br(B \rightarrow X_s \gamma) = (3.32\pm0.15)\times10^{-4}$~\cite{Amhis:2019ckw}, $Br(B_s \rightarrow \mu^{+}\mu^{-}) = (2.69^{+0.37}_{-0.35})\times10^{-9}$~\cite{LHCb:2020zud,Aaij:2021nyr}, $Br(B^+ \rightarrow \tau^{+}\nu) = (1.06\pm0.19)\times10^{-4}$~\cite{Amhis:2019ckw}. We use \texttt{micrOMEGAs-5.0.8}~\cite{B_langer_2002,Belanger:2005kh,Belanger:2018ccd} to compute the corresponding branching ratios for points in our allowed parameter space, and require them to fall within $2\sigma$ uncertainty of the current best-fit values. We also include a theoretical uncertainty of $10\%$ while computing the $1\sigma$ ranges. Constraints on $\Delta M_{D}$, $\Delta M_{S}$, the mass differences between $B_d^0, \bar B_d^0$ and $B_s^0, \bar B_s^0$ respectively, are also imposed through the \texttt{NMSSMTools-5.5.3} package.

\item {\bf Constraints from LHC searches:} 
The composition of heavy Higgs bosons $H_{2}$ and $A_{1}$ in the parameter space of interest are presented in Fig.~\ref{fig:heavy_higgs_composition}~(left) and (right), respectively. In Fig.~\ref{fig:heavy_higgs_composition}, $S_{21}^{2} + S_{22}^{2}$ and $S_{23}^{2}$ represents the doublet and singlet content in $H_{2}$. Similarly, the doublet and singlet admixture in $A_{1}$ is denoted by $P_{11}^{2}$ and $P_{12}^{2}$, respectively. We observe that both $H_{2}$ and $A_{1}$ have a dominant singlino composition~($\gtrsim 90\%$) leading to immunity from heavy Higgs search limits. We would like to note that the heaviest neutral Higgses $H_{3}$ and $A_{2}$ have a dominant doublet composition and have masses above $\gtrsim 2~\mathrm{TeV}$, thereby, remaining outside the direct reach of current LHC. 
\begin{figure*}[!htb]
	\centering
	\includegraphics[height=5.2cm,width=7cm]{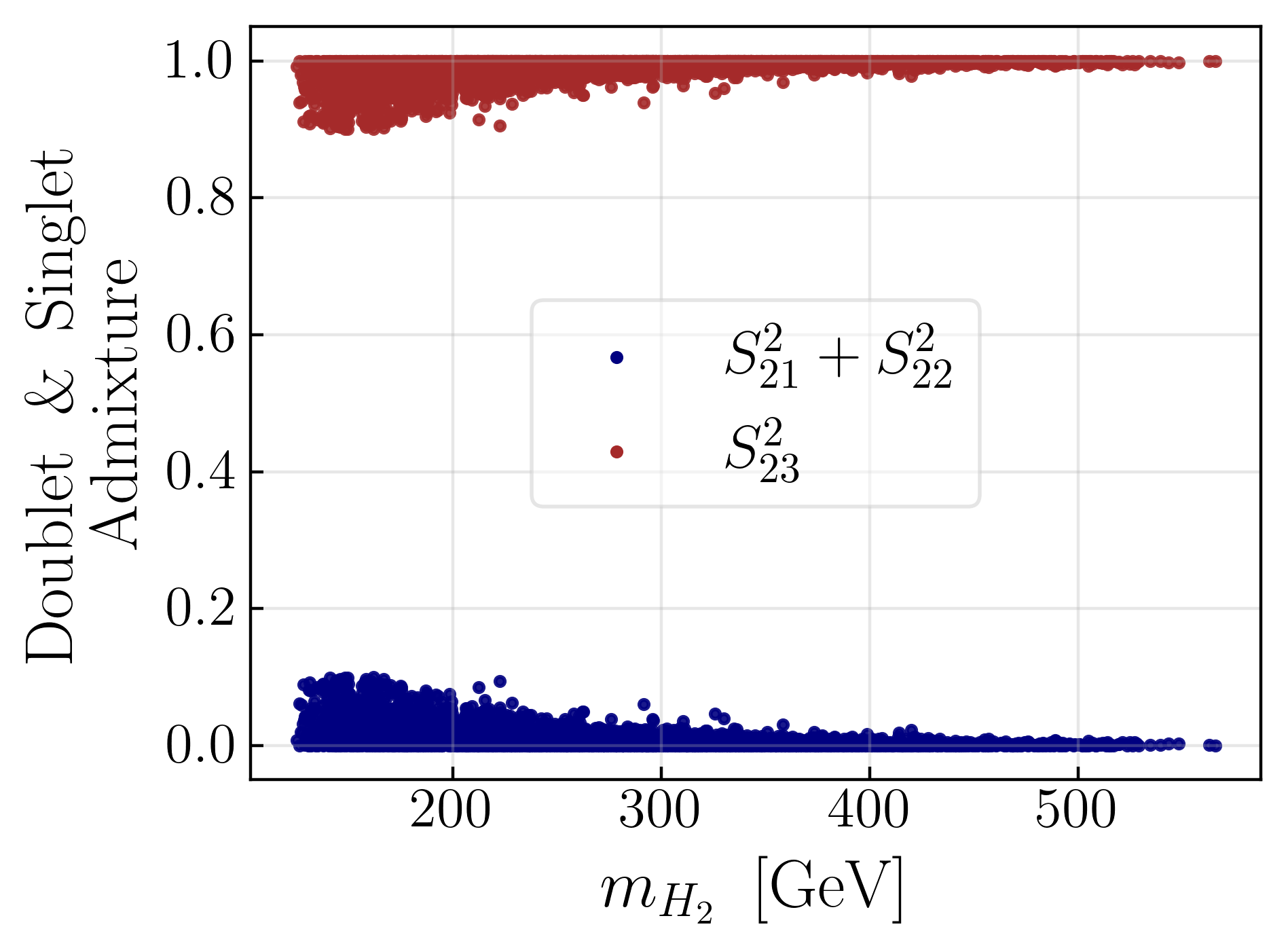}\hspace{1cm} \includegraphics[height=5.2cm,width=7cm]{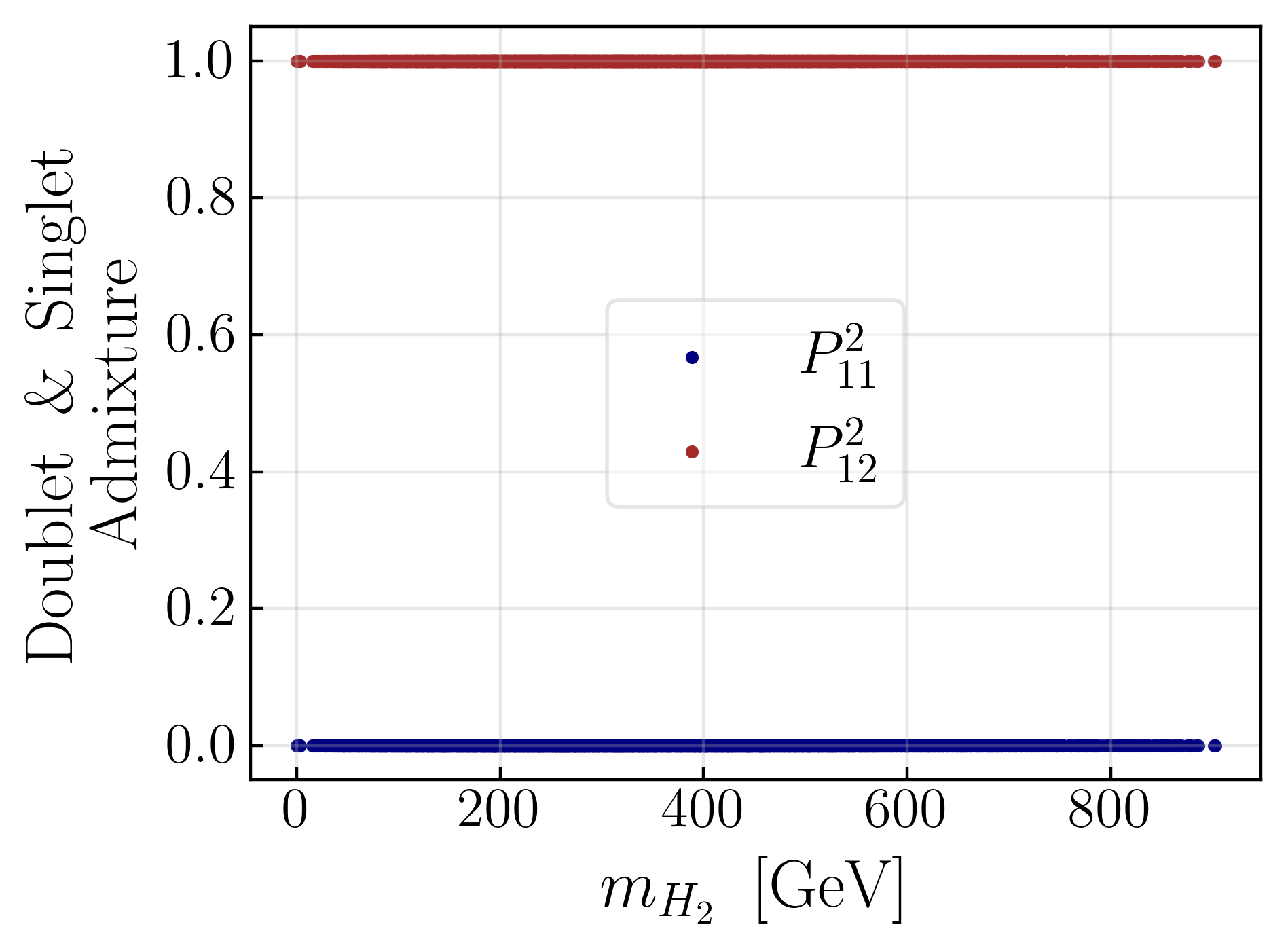}
	\caption{Singlet and doublet admixtures in $H_{2}$~(left panel) and $A_{1}$~(right panel) for parameter space allowed by light Higgs mass constraints, LEP limits, Higgs signal strength constraints and bounds from flavor physics. $S_{21}^{2} + S_{22}^{2}$ and $S_{23}^{2}$ corresponds to the doublet and singlet admixtures, respectively, in $H_{2}$. $P_{11}^{2}$ and $P_{12}^{2}$ represent the doublet and singlet admixture, respectively, in $A_{1}$.}
	\label{fig:heavy_higgs_composition}
\end{figure*}

Searches for pair produced electroweakinos in the hadronic final state by the ATLAS collaboration using LHC Run-II data collected at $\mathcal{L}=139~\mathrm{fb^{-1}}$ have excluded winos~(higgsinos) up to a mass of $1060~$GeV~(900~GeV) given $m_{\lspone} \leq 400~$GeV~(240~GeV) and the mass difference between the decaying wino~(higgsino) and the LSP is larger than 400~GeV~(450~GeV) at $95\%$ CL~\cite{PhysRevD.104.112010}. However, these searches assume a simplified framework where the heavier wino/higgsino-like electroweakinos $\lspi$ directly decays into the LSP $\lspone$ with $Br(\lspi \to \lspone+Z)+Br(\lspi \to \lspone + h_{125}) = 100\%$. Let us analyze the implications of these constraints on  the parameter space considered in this work. Within the parameter space of our interest, $\lspthree, \lspfour$ and $\chonepm$ have dominant higgsino composition with masses ranging from $\sim$500~GeV to 2~TeV, while the wino-like $\lspfive$ and $\chtwopm$ are decoupled $m_{\lspfive/\chtwopm} \gtrsim 2~\mathrm{TeV}$. In principle, $\lspthree, \lspfour$ and $\chonepm$ have two potential pathways for decay, either through the intermediate bino-like $\lsptwo$ or directly into the singlino-like $\lspone$. The partial decay width for $\lspthree/\lspfour \to \lspone Z^{*}$ is determined by the higgsino admixture in $\lspone$ which is directly proportional to $\lambda$. The partial decay width for $\lspthree/\lspfour \to \lspone H$  also has a similar $\lambda$-dependence by virtue of the second term in the NMSSM superpotential in Eq.~(\ref{eq:NMSSM_superpotential}). Therefore, the partial decay widths for both channels through which $\lspthree/\lspfour$ can directly decay into $\lspone$ are $\mathcal{O} (\lambda^{2})$~\cite{Ellwanger:2018zxt}. Thus, they are far smaller relative to the partial decay widths for $\lspthree/\lspfour$ decaying into the bino-like $\lsptwo$. Similar arguments can also be extended to the higgsino-like $\chonepm$. Consequently, in the present scenario, directly produced $pp \to \lspthree\chonepm + \lspfour\chonepm$ will dominantly undergo cascade decay via $(\lspthree/\lspfour \to (\lsptwo \to \lspone Z^{(*)}/H^{(*)}) Z/H)(\chonepm \to (\lsptwo \to \lspone Z^{(*)}/H^{(*)}) W^{\pm})$ leading to final states that are markedly different from those considered in the ATLAS search~\cite{PhysRevD.104.112010}. Furthermore, the allowed points in the parameter space with $m_{\lspthree,\lspfour,\chonepm} \lesssim 1~\mathrm{TeV}$ and a dominant higgsino admixture in $\lspthree,\lspfour$ and $\chonepm$~($\gtrsim 90\%$), correspond to mass differences between $\{\lspthree/\lspfour/\chonepm\}$ and $\lsptwo$, which are very close to the ATLAS search threshold~$\sim 400~\mathrm{GeV}$~\cite{PhysRevD.104.112010}, leading to low efficiencies. Overall, the parameter space of our interest is rather weakly constrained by all the direct electroweakino searches at the LHC.

\item {\bf Constraints from direct detection: }
We apply the most recent upper limits on SI WIMP-nucleon interaction cross-section $\sigma_{SI}$ from Xenon-1T~\cite{XENON:2018rxp} and PandaX~\cite{PandaX-4T:2021bab}, on SD WIMP-proton interaction cross-section $\sigma_{SD_{p}}$ from PICO-60~\cite{PhysRevD.100.022001} and SD WIMP-neutron interaction cross-section $\sigma_{SD_{n}}$ from Xenon-1T~\cite{XENON:2019rxp}. We impose these direct detection~(DD) limits after all the constraints discussed till now and find that these direct detection searches do not lead to any additional constraints on the parameter space. In fact, the predicted SI DD cross-sections fall below the neutrino floor for the entirety of the currently allowed points in the scanned parameter space. Hence, these would be inaccessible to future DD experiments based on $\sigma_{SI}$ measurements. We also examine the projected sensitivity at the future $\sigma_{SD}$ based experiments. For the range of $m_{\lspone}$ in the parameter space of our interest, the most stringent projected sensitivities for $\sigma_{SD_{p}}$ and $\sigma_{SD_{n}}$ are furnished by PICO-250~\cite{Cushman:2013zza} and LZ~\cite{LUX:2016sci}, respectively. However, we observe that neither of them would be sensitive to any of the currently allowed points in the parameter space.

Having discussed the implications of the relevant current constraints, we move on to discuss the features of the currently allowed parameter space in the next section.

\end{itemize}

\begin{figure}[!t]
\centering
\includegraphics[scale=0.52]{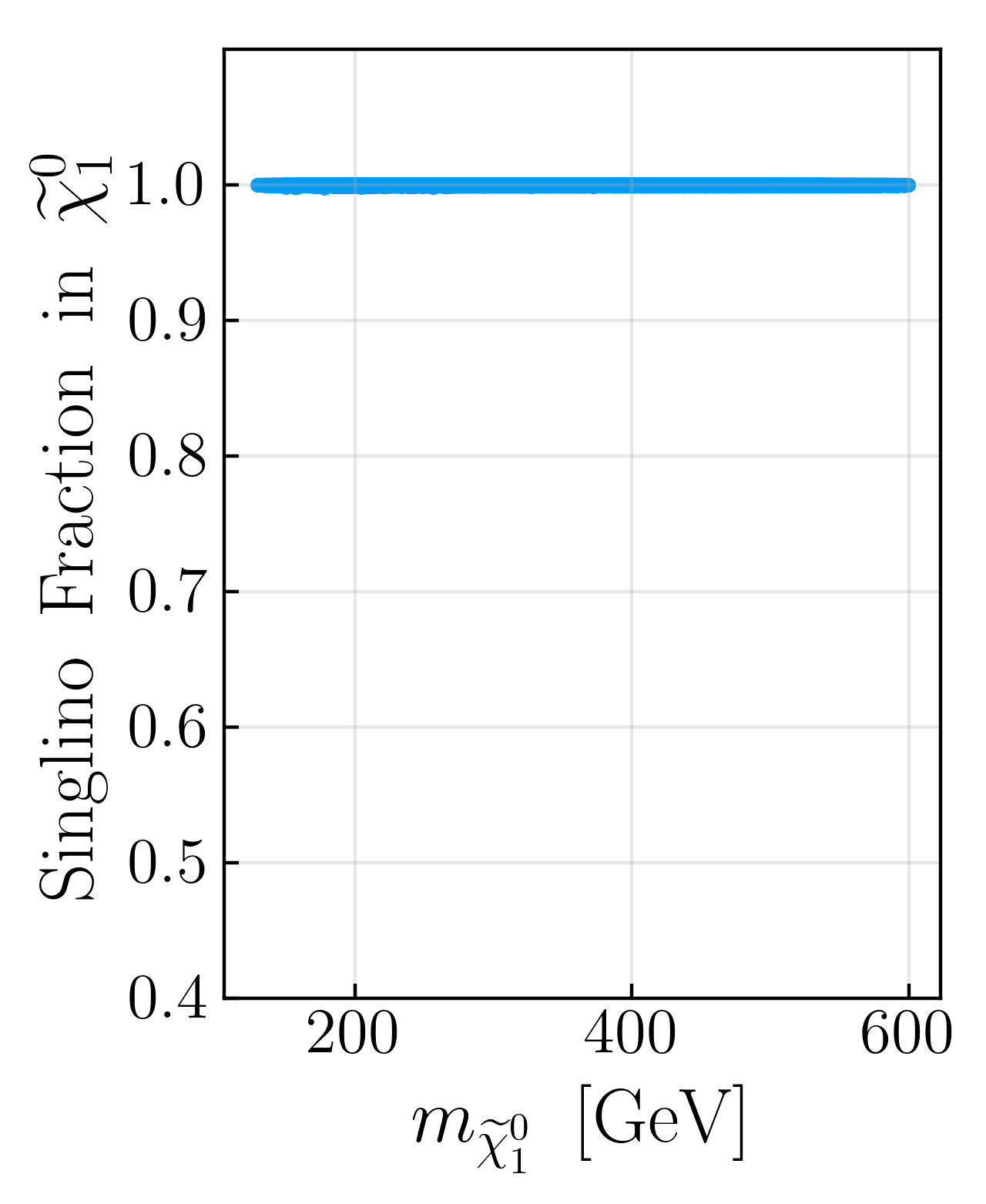}\hspace{1.0cm}\includegraphics[scale=0.52]{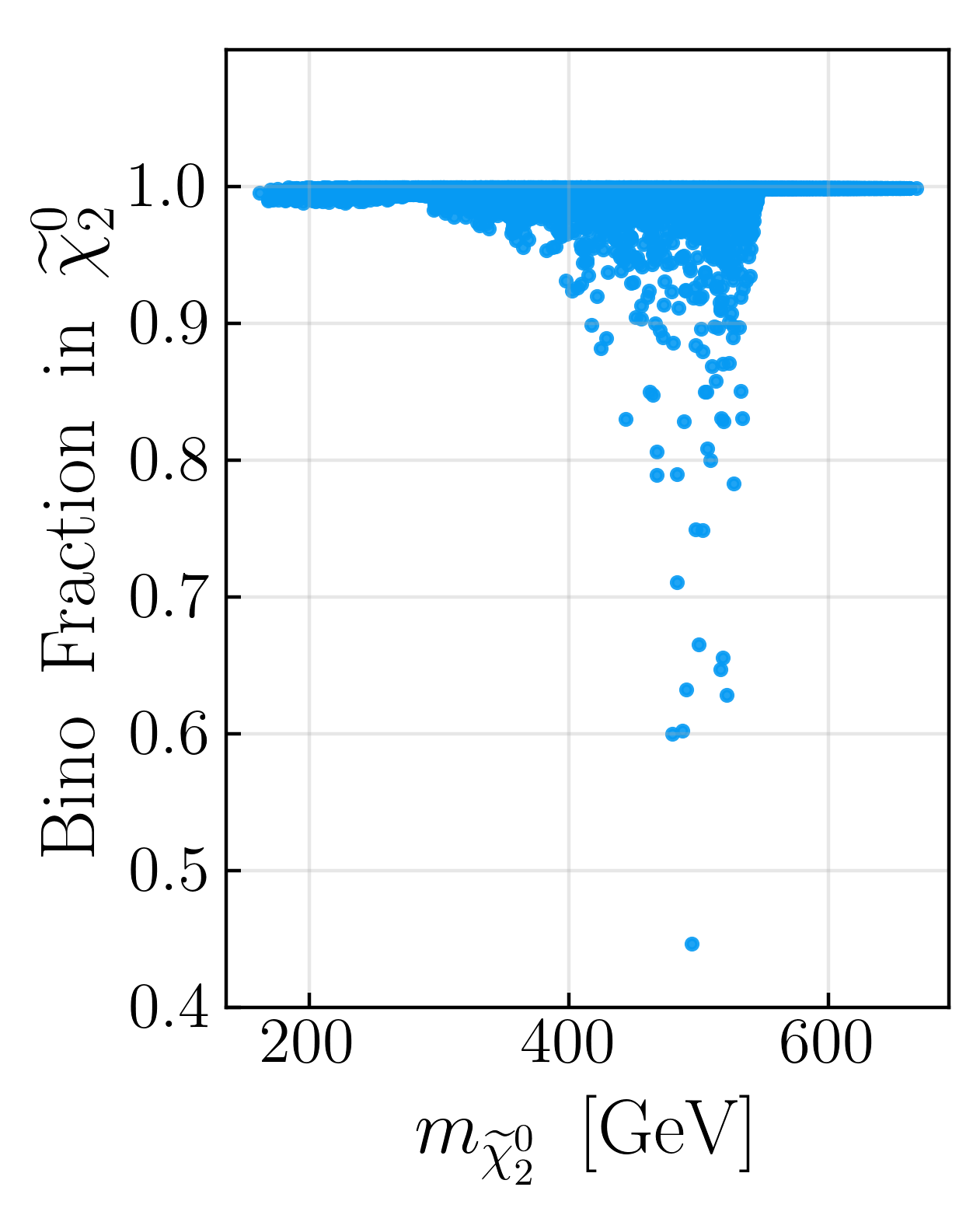}\\
\includegraphics[scale=0.52]{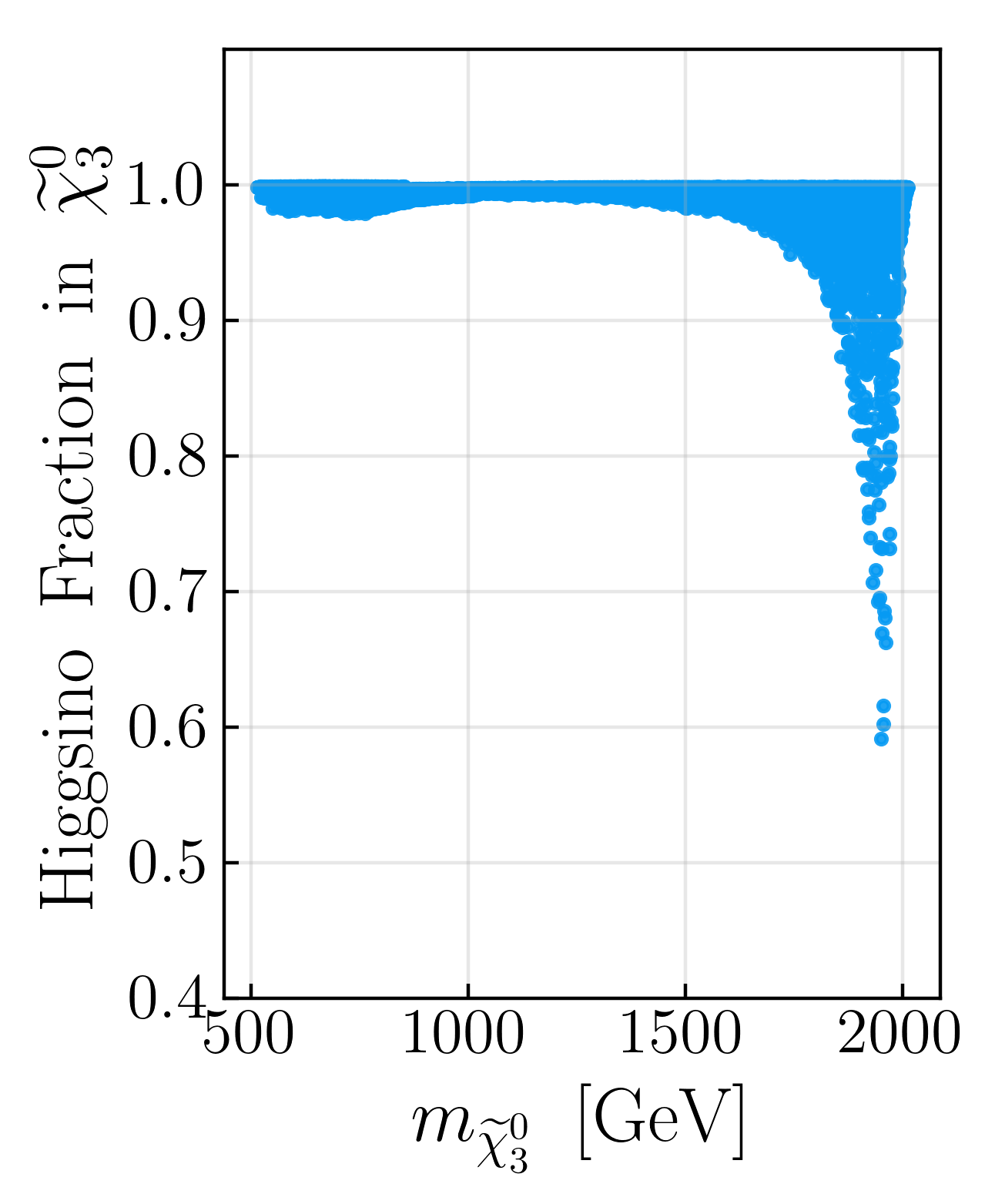}\hspace{1.0cm}\includegraphics[scale=0.52]{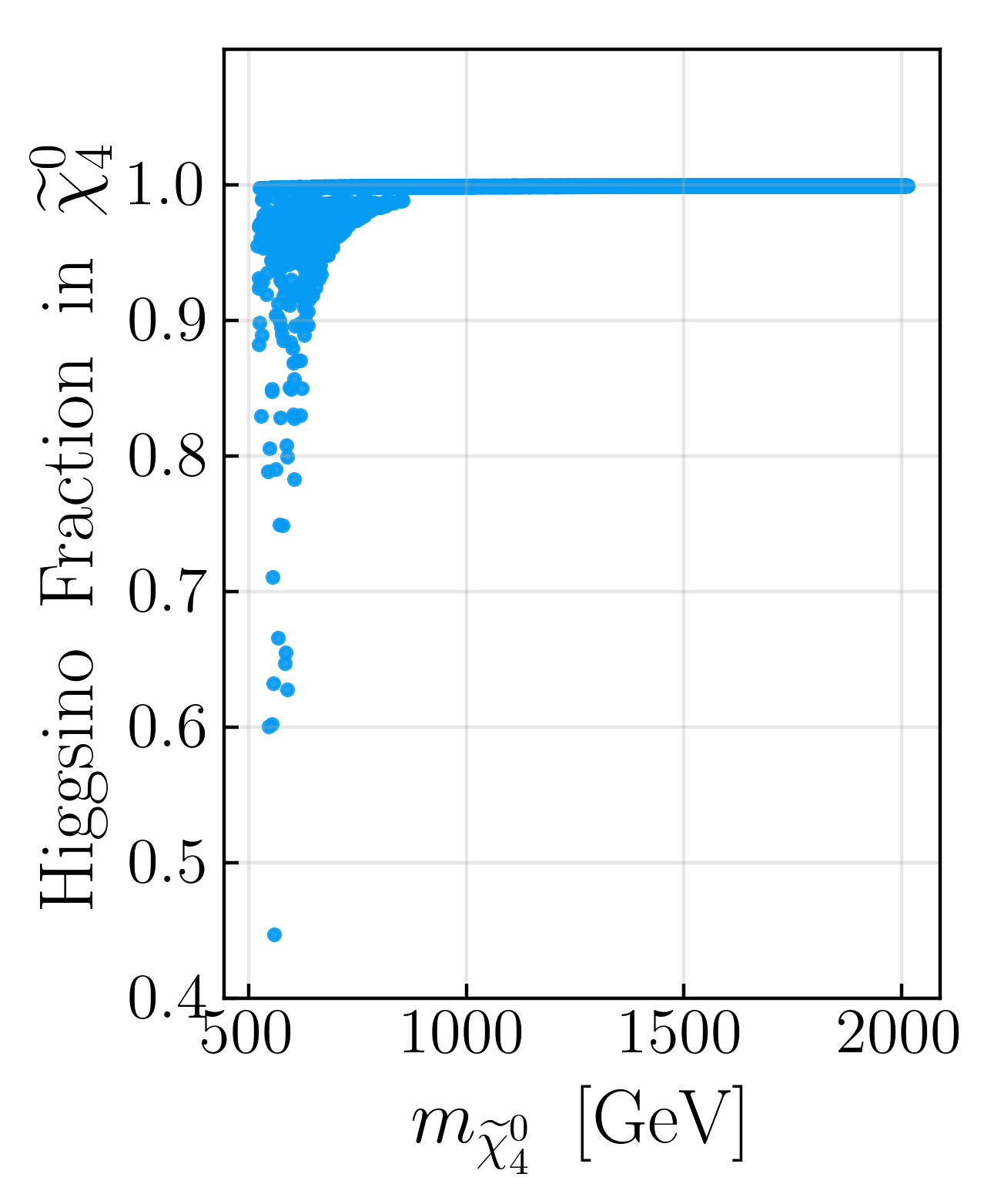}
\caption{Singlino content $N_{11}^2$ in $\lspone$, bino content $N_{21}^2$ in $\lsptwo$, higgsino content ($N_{33}^3 + N_{34}^2$) in $\lspthree$ \& higgsino content ($N_{43}^2 + N_{44}^2$) in $\lspfour$ is shown for currently allowed parameter space.}
\label{fig:neutralino_composition}
\end{figure}

\section{Features of the allowed parameter space}
\label{sec:allowed_parameter_space}

In this section, we examine the features of the allowed parameter space. We would like to emphasize again that our objective is to delineate the  NMSSM parameter space that contains a long-lived bino-like $\lsptwo$ with mass $\sim \mathcal{O}(100)~\mathrm{GeV}$ and is also allowed by the current experimental constraints. In Fig.~\ref{fig:neutralino_composition}, we present the fraction of singlino content in $\lspone$~(upper-left), bino content in $\lsptwo$~(upper-right) and higgsino contents in $\lspthree$~(lower-left) and $\lspfour$~(lower-right) for the allowed points. We observe that the singlino admixture in $\lspone$ is $\gtrsim 99\%$ while $\lsptwo$ has a dominant bino content. Similarly, the heavier neutralinos $\lspthree$ and $\lspfour$ have a dominant higgsino composition.
\begin{figure*}[!t]
\centering
\includegraphics[height=5.2cm,width=7cm]{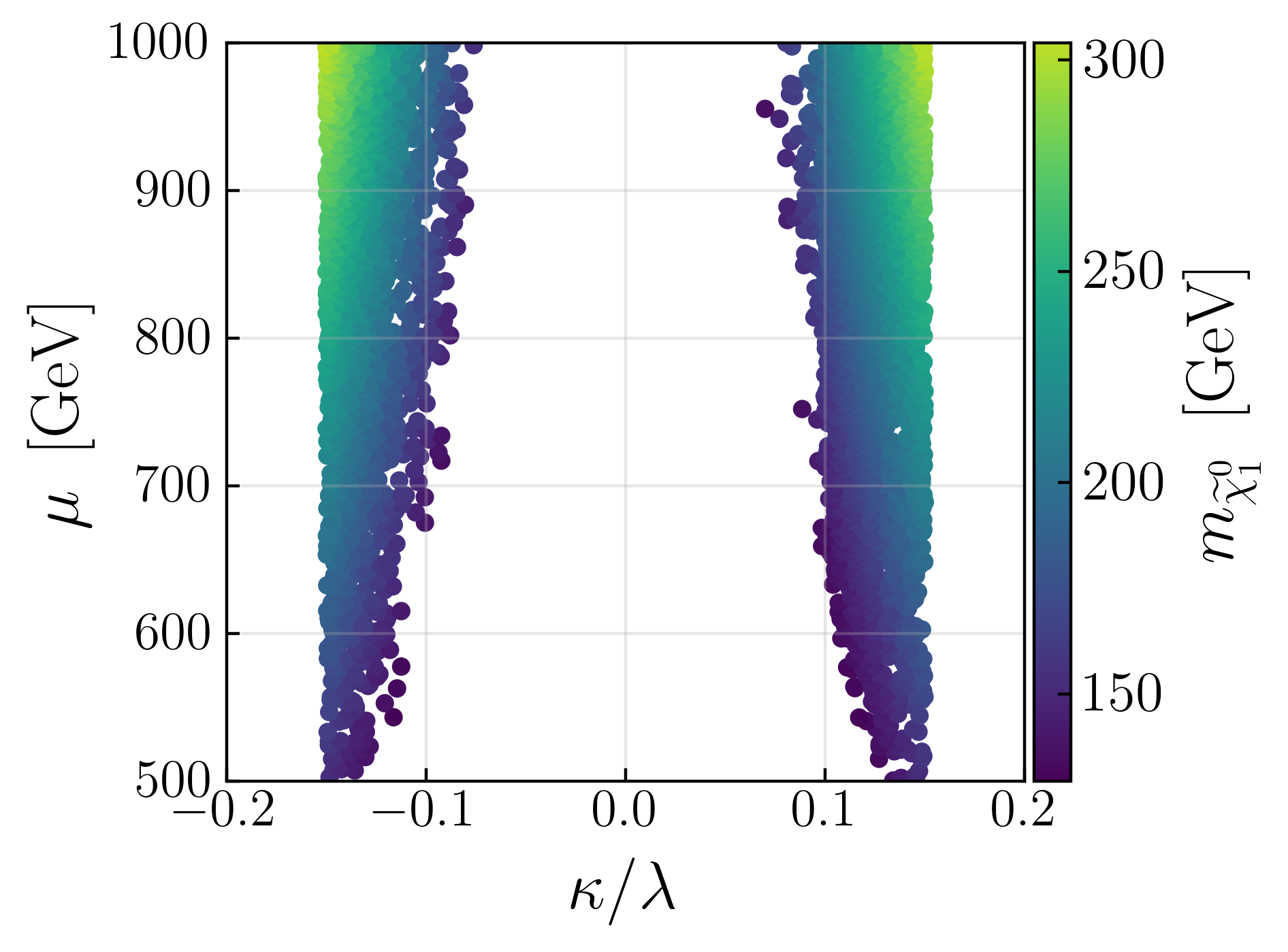}\hspace{1 cm}
\includegraphics[height=5.2cm,width=7cm]{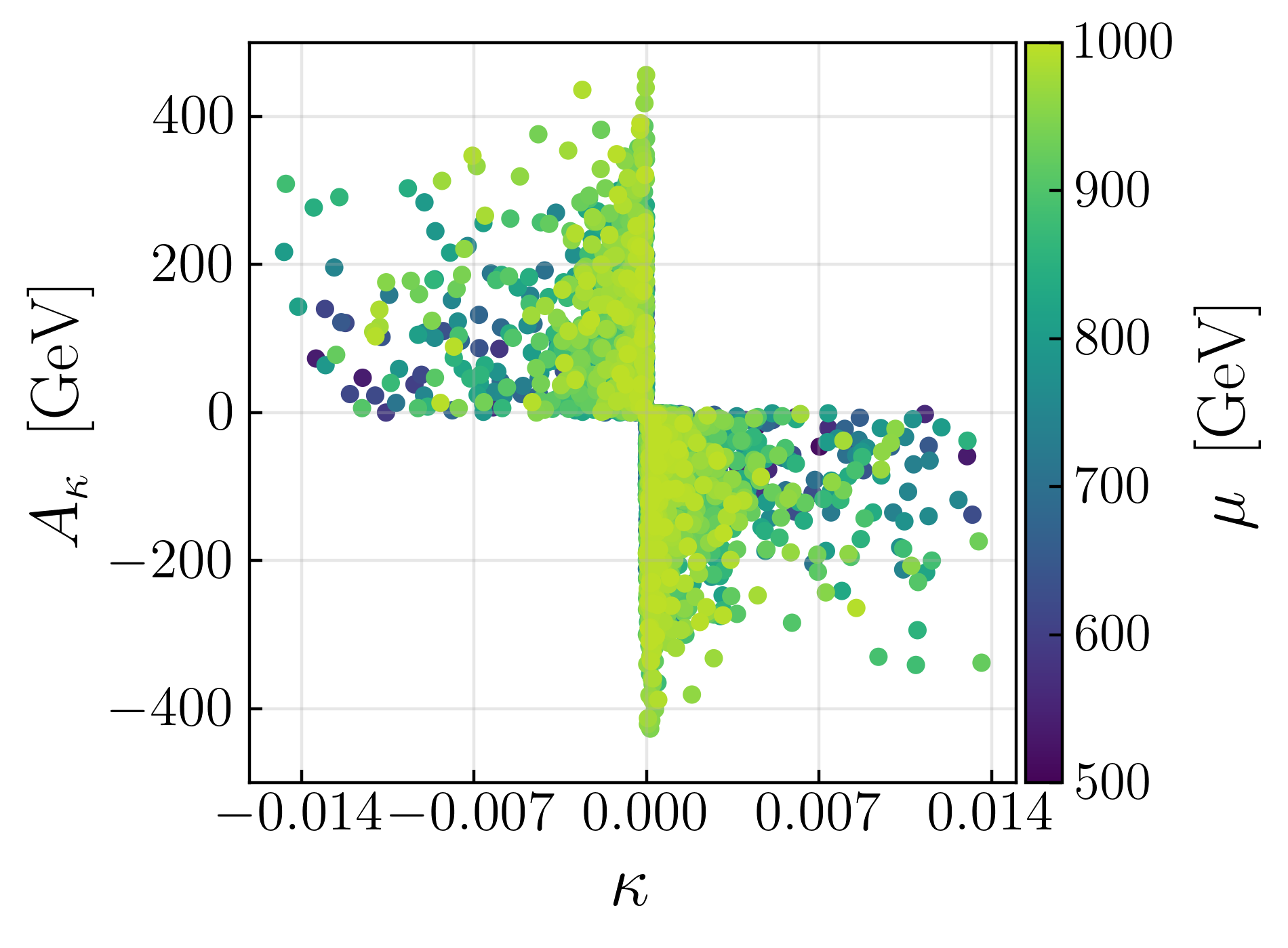}
\label{fig:input_param_2}
\caption{\textit{Left panel:} Allowed parameter space points in the plane of $\kappa/\lambda$ vs. $\mu$. The color palette represents the mass of the singlino-like LSP neutralino. \textit{Right panel:} Allowed parameter space in the $k$ vs. $A_{\kappa}$ plane. The color palette represents the higgsino mass parameter $\mu$.}
\label{fig:param_feature_2}
\end{figure*}

As noted in Eq.~(\ref{eq:param_scan}), we consider $|\kappa|/\lambda < 0.15$ besides $500~\mathrm{GeV}\leq \mu \leq 1~\mathrm{TeV}$ and $2~\mathrm{TeV} \leq M_{2} \leq 3~\mathrm{TeV}$ in order to obtain a dominantly singlino-like LSP. We illustrate the allowed points in the $k/\lambda$ vs. $\mu$ plane in the left panel of Fig.~\ref{fig:param_feature_2}. The color palette represents the mass of $\lspone$. We observe that $m_{\lspone}$ increases moderately with $\mu$ for a fixed value of $\kappa/\lambda$. At a given $\kappa/\lambda$, $\kappa$ also increases with $\mu$ since $\mu \sim \lambda v_{S}$. This leads to an increment in the mass of singlino-like $\lspone$ since $m_{\lspone} \sim 2\kappa v_{S}$. For a fixed $\mu$, the points with the smallest values of $|\kappa|/\lambda$ correspond to lowest, $m_{\lspone}$ as implied from Eq.~(\ref{eq:singlino_mass})\footnote{Eq.~(\ref{eq:singlino_mass}) can be adapted to $|\kappa|/\lambda \sim 0.5 \times m_{\lspone}/\mu$, which shows a direct correlation between $m_{\lspone}$ and $|\kappa|/\lambda$ at a given $\mu$.}. Considering $\mu \sim 1~\mathrm{TeV}$ and our assumption of $|\kappa|/\lambda < 0.15$, $m_{\lspone}$ is restricted to $m_{\lspone} \lesssim 300~\mathrm{GeV}$, which is consistent with the observations in Fig.~\ref{fig:param_feature_2}~(left). Similarly, at $\mu \sim 1~\mathrm{TeV}$, the lowest value of $m_{\lspone}$, $m_{\lspone} \sim 125~\mathrm{GeV}$, implies a lower limit of $|\kappa|/\lambda \gtrsim 0.063$. As we move towards smaller values of $\mu$, the lower limit on $|\kappa|/\lambda$ gets stronger, for instance at $\mu \sim 500~\mathrm{GeV}$, we obtain $|\kappa|/\lambda \gtrsim 0.125$ for $m_{\lspone} \sim 125~\mathrm{GeV}$, as also observed in Fig.~\ref{fig:param_feature_2}~(left). We also present the allowed points in the $\kappa$ vs $A_{\kappa}$ plane in the right panel of Fig.~\ref{fig:param_feature_2}. While both $\kappa$ and $A_{\kappa}$ can take positive or negative values, their product is required to be $\lesssim 0$. This requirement is implied by Eq.~(\ref{eqn:pseudoscalar_mass}) where a positive value of $M_{P,22}^{2}$ at small $\lambda$ is guaranteed only if $\kappa A_{\kappa} < 0$. 
\begin{figure}[!t]	
	\centering
	\includegraphics[height=5.2cm,width=7cm]{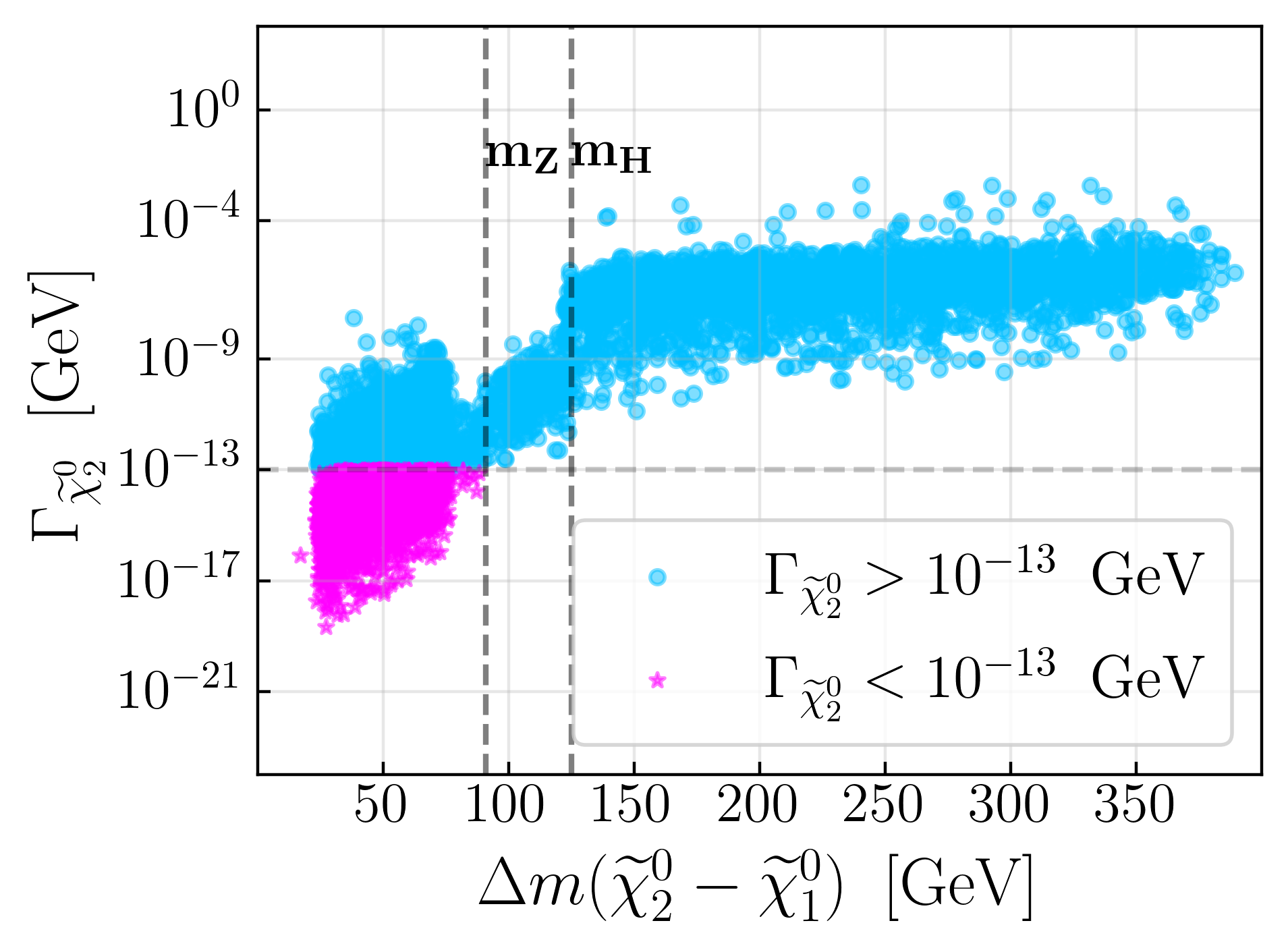}
	\caption{Allowed parameter space are presented in the plane of mass difference between $\lsptwo$ and $\lspone$ $\Delta m(\lsptwo - \lspone)$ vs. the decay width of $\lsptwo$ $\Gamma_{\lsptwo}$. The vertical black dashed lines represent the $Z$ and $H_{125}$ on-shell conditions. Parameter space with $\Gamma_{\lsptwo} \leq 10^{-13}~\mathrm{GeV}$ and $\Gamma_{\lsptwo} > 10^{-13}~\mathrm{GeV}$ are illustrated in pink and cyan colors, respectively.}
	\label{fig:DW_chi2}
\end{figure}
One of the most exciting aspects of the parameter space is the presence of long-lived neutralino. In Fig.~\ref{fig:DW_chi2}, we present the decay width of $\lsptwo$ ($\Gamma_{\lsptwo}$) as a function of the mass difference between $\lsptwo$ and $\lspone$, $\Delta m(\lsptwo - \lspone)$. We concentrate on the region highlighted by pink with $\Gamma_{\lsptwo} \leq 10^{-13}$. A decay width of $\Gamma \sim 10^{-13}~\rm{GeV}$ roughly translates to $c\tau \sim \mathcal{O}(0.1)~\rm{mm}$. We observe that $\Gamma_{\lsptwo}$ can be smaller than $\lesssim 10^{-13}~\mathrm{GeV}$ when $\Delta m(\lsptwo - \lspone) < m_{Z}$. In this region, only three body decays of $\lsptwo$ are viable \cite{Das:2011dg}. 
We observe that $\Gamma_{\lsptwo}$ can be as small as $\sim 10^{-19}~\mathrm{GeV}$ for relatively smaller mass differences $\Delta m(\lsptwo - \lspone) \lesssim 40~\mathrm{GeV}$. Most notably, this mass difference is still large enough to produce energetic final states as $\lsptwo$ decays. Such configurations are not possible in MSSM with neutralino LSP and are a unique feature of NMSSM scenario we consider. 

\begin{figure*}[!t]
    \centering
    \includegraphics[scale = 0.41]{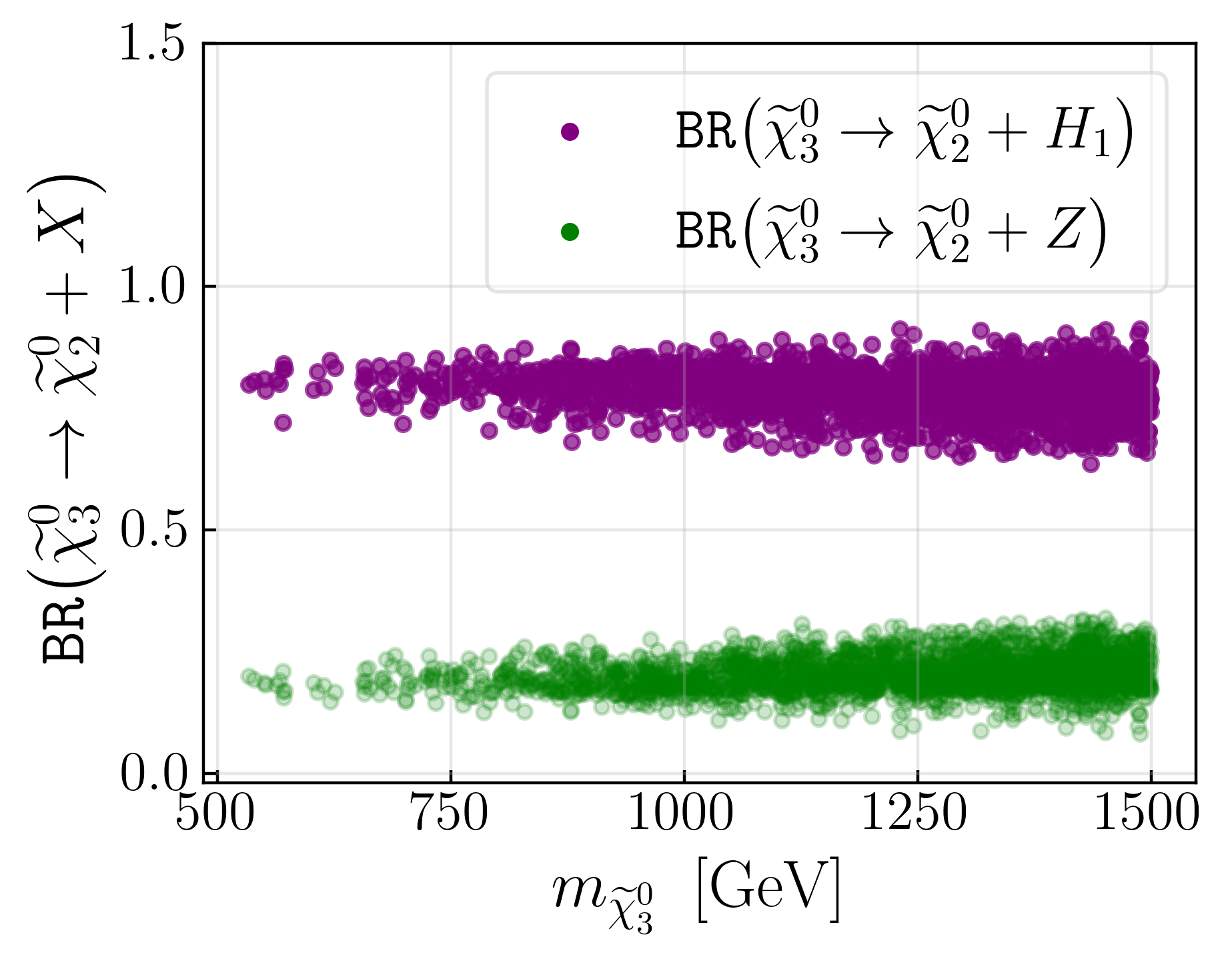}
    \includegraphics[scale = 0.41]{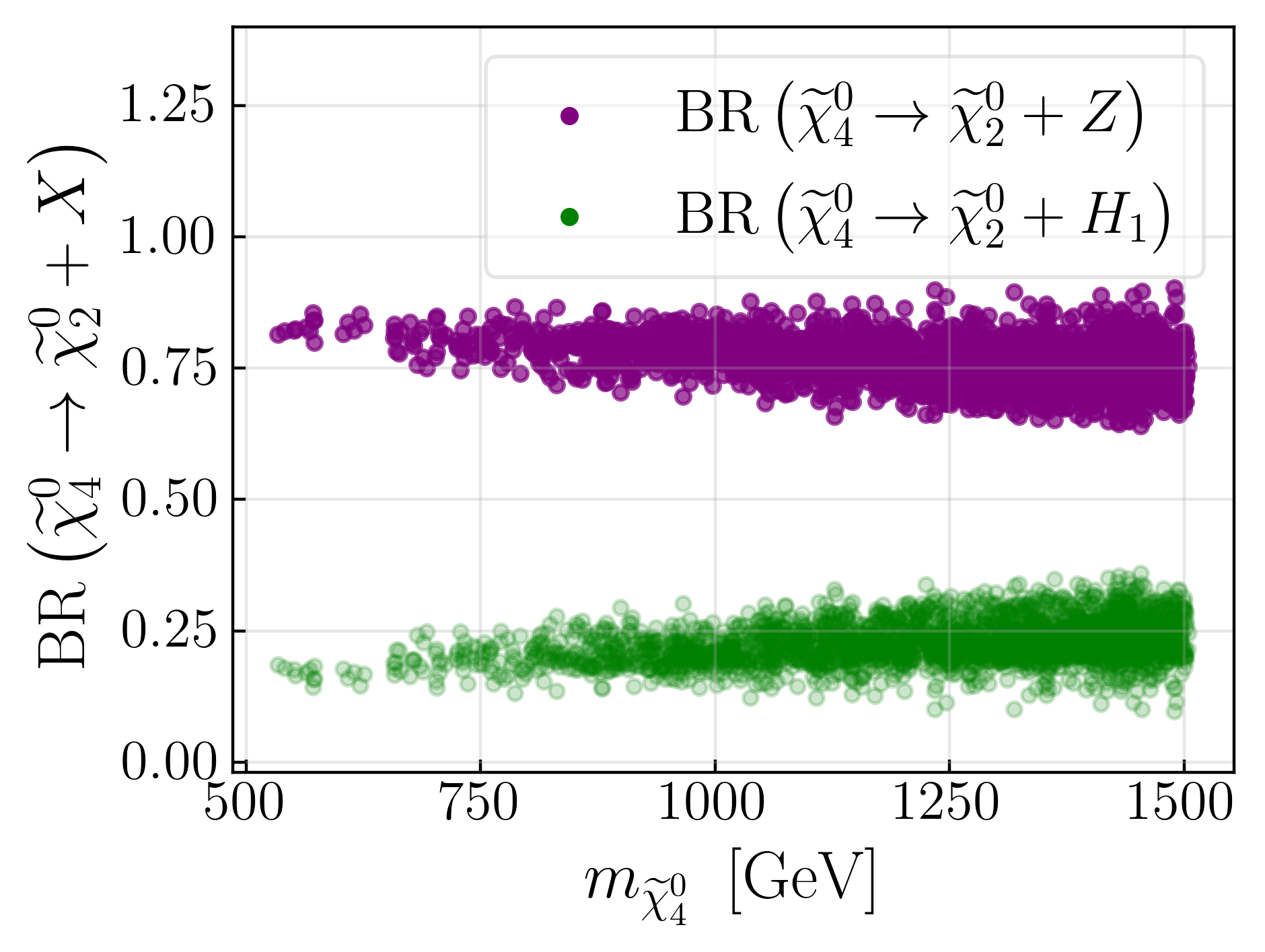}
    \includegraphics[scale = 0.41]{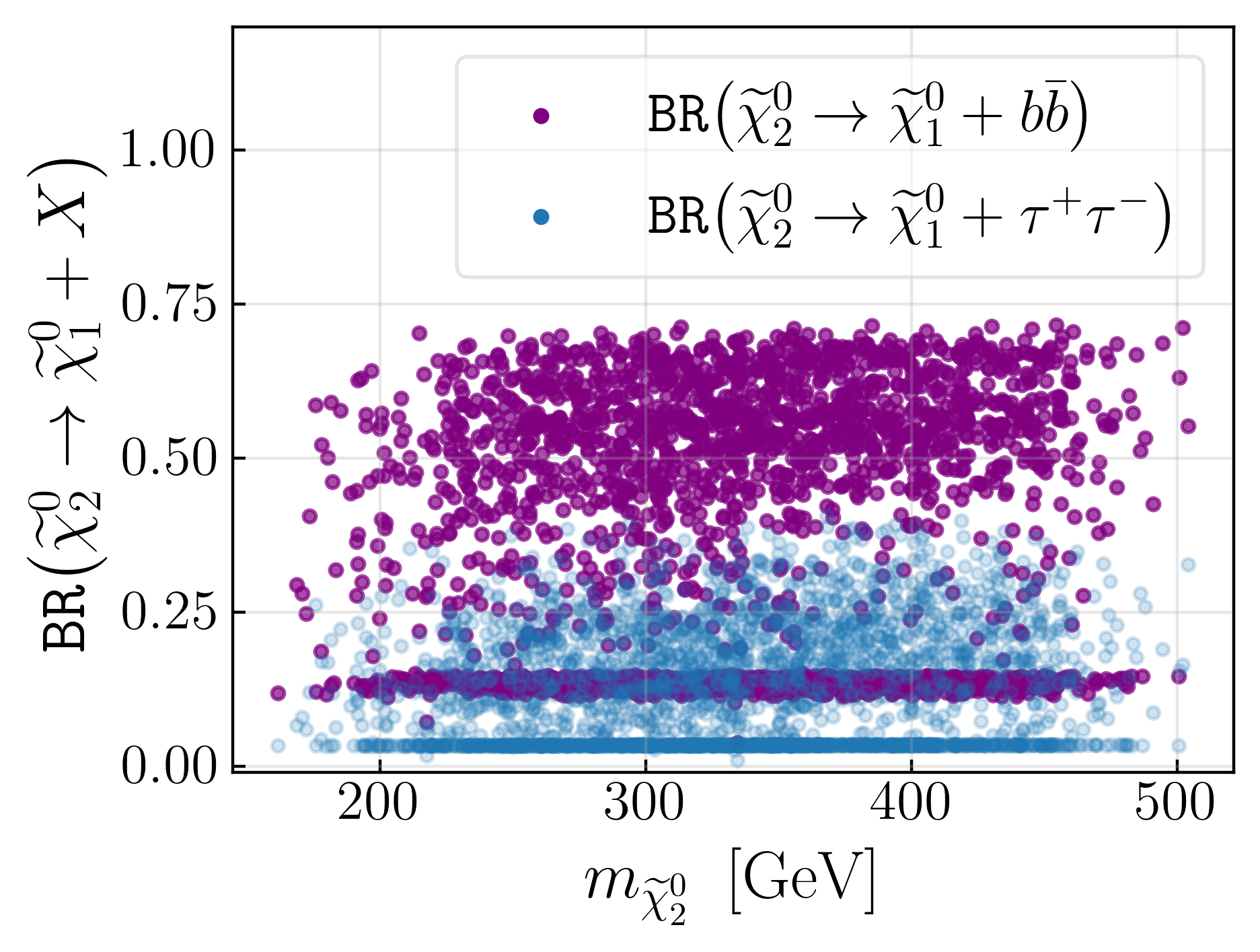}
    \caption{\small \textit{Left and Central panels:} Branching fractions for the dominant decay modes of $\lspthree$~(\textit{left panel}) and $\lspfour$~(\textit{central panel}) are shown for allowed parameter space. The red and purple colored points represent the branching ratio of $\lspthree/\lspfour$ into $\lsptwo + Z$ and $\lsptwo + H_{1}$ final states. \textit{Right panel:} Branching fraction for $\lsptwo$ are shown for the allowed parameter space. The red and purple points denote the branching fractions $Br(\lsptwo \to \lspone + b\bar{b})$ and $Br(\lsptwo \to \lspone + \tau^{+}\tau^{-})$, respectively.}
    \label{fig:br_chi3_chi4}
\end{figure*}

Thus, the allowed points can have long-lived bino-like $\lsptwo$ with decay widths up to $\sim 10^{-19}~$GeV. Furthermore, the heavier neutralinos $\lspthree,\lspfour$ and the lightest chargino $\chonepm$ have a dominant higgsino admixture while $\tilde{\chi}_{5}^{0}$ and $\chtwopm$ are wino-like. We have set $M_{2}$ to be above $\gtrsim 2~$TeV, thus, decoupling $\tilde{\chi}_{5}^{0}$ and $\chtwopm$ from the rest of the electroweakinos. Since the LSP $\lspone$ has a dominant singlino content, the higgsino-like $\lspthree, \lspfour$ mostly decays through the intermediate bino-like $\lsptwo$, $\lspthree/\lspfour \to \lsptwo + H_{1}/Z$, instead of decaying directly into $\lspone + X$ states. In the left and central panels of Fig.~\ref{fig:br_chi3_chi4}, we present the branching ratios $Br(\lspthree \to \lsptwo + H_{1}/Z)$ and $Br(\lspfour \to \lsptwo + H_{1}/Z)$ for the allowed parameter space. We observe that $\lspthree$~($\lspfour$) can decay via $\lsptwo + H_{1}$~($\lsptwo + Z$) with branching fractions as large as $\sim 0.9$. Interestingly, the $\lsptwo$'s can further undergo three body decay $viz$ $\lspone + b\bar{b}/\tau^{+}\tau^{-}/jj$~($j=u,c,d,s$), when $\Delta m(\lsptwo - \lspone) \lesssim m_{Z}$. We illustrate the branching ratios of $\lsptwo \to \lspone b \bar{b}$ and $\lsptwo \to \lspone \tau^{+} \tau^{-}$ for allowed parameter space with $\Delta m(\lsptwo - \lspone) \lesssim m_{Z}$ in Fig.~\ref{fig:br_chi3_chi4}~(right panel). The figure shows array of points in the region of branching ratio $ < 0.2$ , for both the $b\bar{b} ~\&~ {\rm the}~\tau^+\tau^-$ decay modes. These unusual points have $\kappa < 0 $. The change in the sign of the term containing $\kappa$ in the mass matrix causes a change in the mixing patterns and also the mass eigenvalues, increasing admixture of bino by in the LSP by $\mathcal{O}(2)$ than the rest. Hence, the $\lsptwo$ seems to be decaying mainly through $Z^*$ instead of off-shell Higgs mediation. This is also substantiated from the observed universality of the leptonic decay modes of these points, $Br(\lsptwo \to \lspone e^{+}e^{-}) = Br(\lsptwo \to \lspone \mu^{+}\mu^{-}) \approx Br(\lsptwo \to \lspone \tau^{+}\tau^{-})$, for these points.

\section{ Discovery potential of LLP decays at the HL-LHC}
\label{sec:Result}

\subsection{Electroweakino pair production rates at the HL-LHC}

\begin{figure}[!t]	
	\centering
	\includegraphics[scale=0.28]{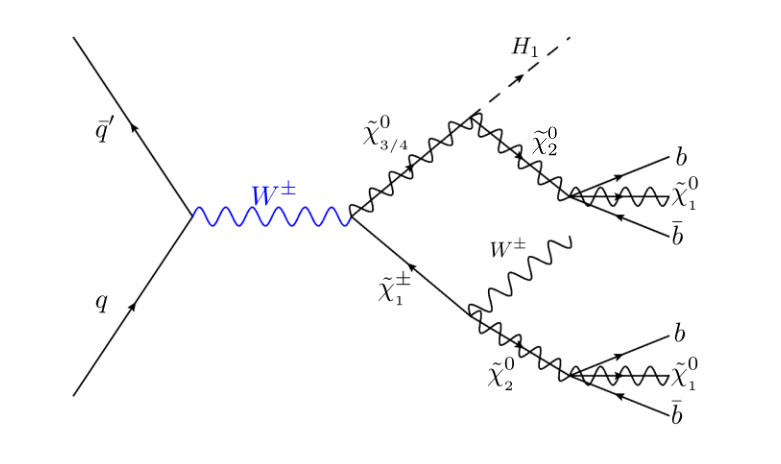}
	\includegraphics[height=5.2cm,width=7cm]{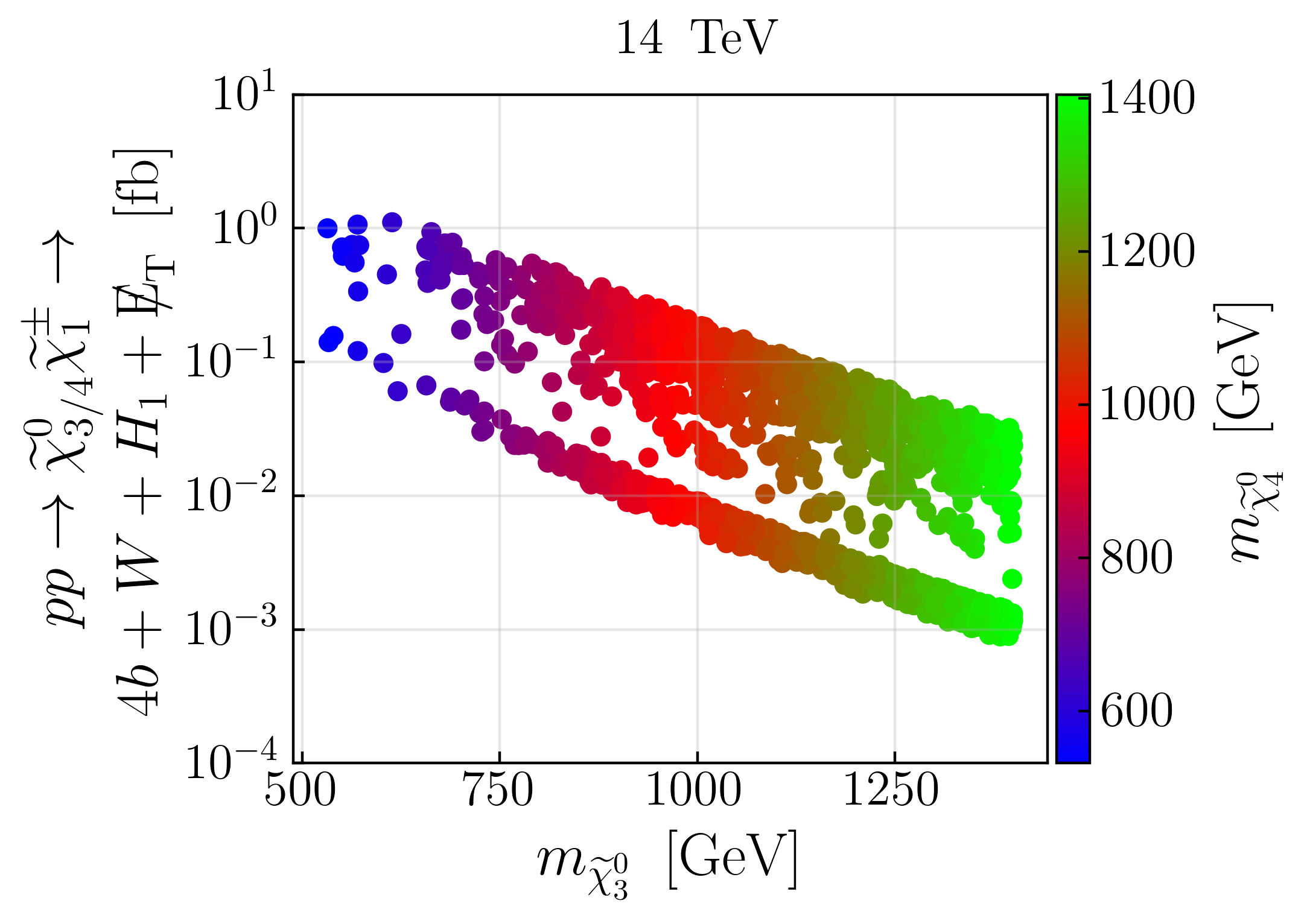}
\caption{\textit{Left panel:} Feynman diagram at leading order for the signal process $\chonepm\lspthree/\lspfour \to \&(\chonepm \to \lsptwo + W^{\pm}, \lsptwo \to \lspone + b\bar{b}) (\lspthree/\lspfour \to \lsptwo + Z/H_{1}, \lsptwo \to \lspone + b\bar{b})$. \textit{Right panel:} Leading order cross-section for the process $pp \to \chonepm\lspthree/\lspfour \to (\chonepm \to \lsptwo + W^{\pm}, \lsptwo \to \lspone + b\bar{b}) (\lspthree/\lspfour \to \lsptwo + H_{1}, \lsptwo \to \lspone + b\bar{b}) \to 4b + WH_{1} + \met$ at $\sqrt{s}=14~\mathrm{TeV}$ for the allowed parameter space with $\Gamma_{\lsptwo} < 10^{-13}~\mathrm{GeV}$.}   
\label{fig:xs_14TeV}
\end{figure}

 As discussed previously, directly produced chargino-neutralino pairs at the HL-LHC can lead to interesting final state topologies involving long-lived $\lsptwo$ in addition to several promptly decaying candidates. One such typical cascade decay chain can be written as follows,
\begin{equation}
\begin{split}
    pp \to &\chonepm\lspthree + \chonepm\lspfour,\\ 
    & \chonepm \to \lsptwo + W^{\pm}, \lsptwo \to \lspone + b\bar{b}, \\
    & \lspthree/\lspfour \to \lsptwo + Z/H_{1}, \lsptwo \to \lspone + b\bar{b}.
    \label{eq:cascade}
\end{split}
\end{equation}
Since the $\lsptwo$ is long-lived, the final states contain displaced b-jets along with $W+Z/H_{1}+\met$. 
In principle, a final state with displaced $\tau$ jets, light jets, or a combination of them can also be realized. For the sake of simplicity, we restrict ourselves to the scenario where the LLP $\lsptwo$ decays into $\lspone + b\bar{b}$~(see Eq.~(\ref{eq:cascade})). For illustration, we show a typical leading order~(LO) Feynman diagram in Fig.~\ref{fig:xs_14TeV}~(left).

The direct production of a chargino-neutralino pair is a pure electroweak process, and controlled by the $W^{\pm}\chipm\lspj$ coupling. We obtain the production cross-sections $\sigma(pp \to \chonepm \lspthree + \chonepm\lspfour)$ for configurations of our interest by rescaling the NLO MSSM production cross-sections computed using \texttt{Prospino} \cite{Beenakker:1996ed} for a pure higgsino-like $\chipm\lspj$ as follows,

\begin{equation}
\begin{split}
   \left.C^2_{W \chonepm \lspj} \right\vert_{j = 3,4} = &\left\{\left(N_{j3}V_{12} - \sqrt{2}\,N_{j2}V_{11}\right)^2 +  \left( N_{j4}U_{12} + \sqrt{2}\,N_{j2}U_{11}\right)^2 \right\}. \\
    \label{eq:red_coup}
\end{split}
\end{equation}

Here, $N_{j3/j4}$ represents the higgsino component while $N_{j2}$ denotes the wino component in the $j^{th}$ neutralino. The higgsino and wino admixtures in $\chonepm$ are represented by $V_{12}/U_{12}$ and $V_{11}/U_{11}$, respectively. The NMSSM parameter space considered in the present study characterizes a dominant higgsino composition in $\lspthree/\lspfour$ and $\chonepm$. Correspondingly, both $N_{33}^{2} + N_{34}^{2}$ and $N_{43}^{2} + N_{44}^{2}$ $\sim 1$. Similarly, $U_{12}$ and $V_{12}$ are $\sim 1$. On the other hand, $N_{32}, N_{42}, V_{11}$ and $U_{11}$ are $\ll 1$. Therefore, from Eq.~(\ref{eq:red_coup}) $(N_{j3}V_{12})^{2}$ and $(N_{j4}U_{12})^{2}$ are the only relevant terms to compute $\sigma(pp \to \chonepm\lspthree + \chonepm\lspfour)$. The scaled production cross-section is then multiplied by the relevant branching ratios for $\lsptwo$, $H_1/Z$ and $W^{\pm}$. In Fig.~\ref{fig:xs_14TeV}~(right), for all the allowed points featuring a long-lived $\lsptwo$~(Fig.~\ref{fig:DW_chi2}, pink points) we present cross-section $(\sigma(pp \to \chonepm\lspthree + \chonepm\lspfour) \times Br(\chonepm \to \lsptwo + W^{\pm}, \lsptwo \to \lspone + b\bar{b}) \times Br(\lspthree/\lspfour \to \lsptwo + H_{1}, \lsptwo \to \lspone + b\bar{b})$) at $\sqrt{s}=14~\mathrm{TeV}$ as a function of $m_{\lspthree}$ indicating $m_{\lspfour}$ values by different colors. Two bands in the figure with larger (smaller) values are related to two sets of points with different branching fraction to $b\bar{b}$, $>0.2 (<0.2)$ discussed in the previous section.
This cross-section of the entire cascade chain can be as large as $\mathcal{O}(1)~\mathrm{fb}$ and $\mathcal{O}(0.1)~\mathrm{fb}$ at $m_{\lspthree} \sim 500~\mathrm{GeV}$ and $\sim 1~\mathrm{TeV}$, respectively. Considering the large production rates at the HL-LHC, we perform a detailed collider study to explore the projected sensitivity for some benchmark scenarios selected from the allowed parameter space. We focus on direct electroweakino production with the final state containing $WZ/H_{1}$ and displaced $b$ jets with $\met$. Before moving to a discussion of the details of the collider analysis, let us examine some generic features of long-lived particles that are relevant for the present study.

\subsection{Kinematic features of LLP decays}
\begin{figure}[!t]	
	\centering
	\includegraphics[height=5.2cm,width=7cm]{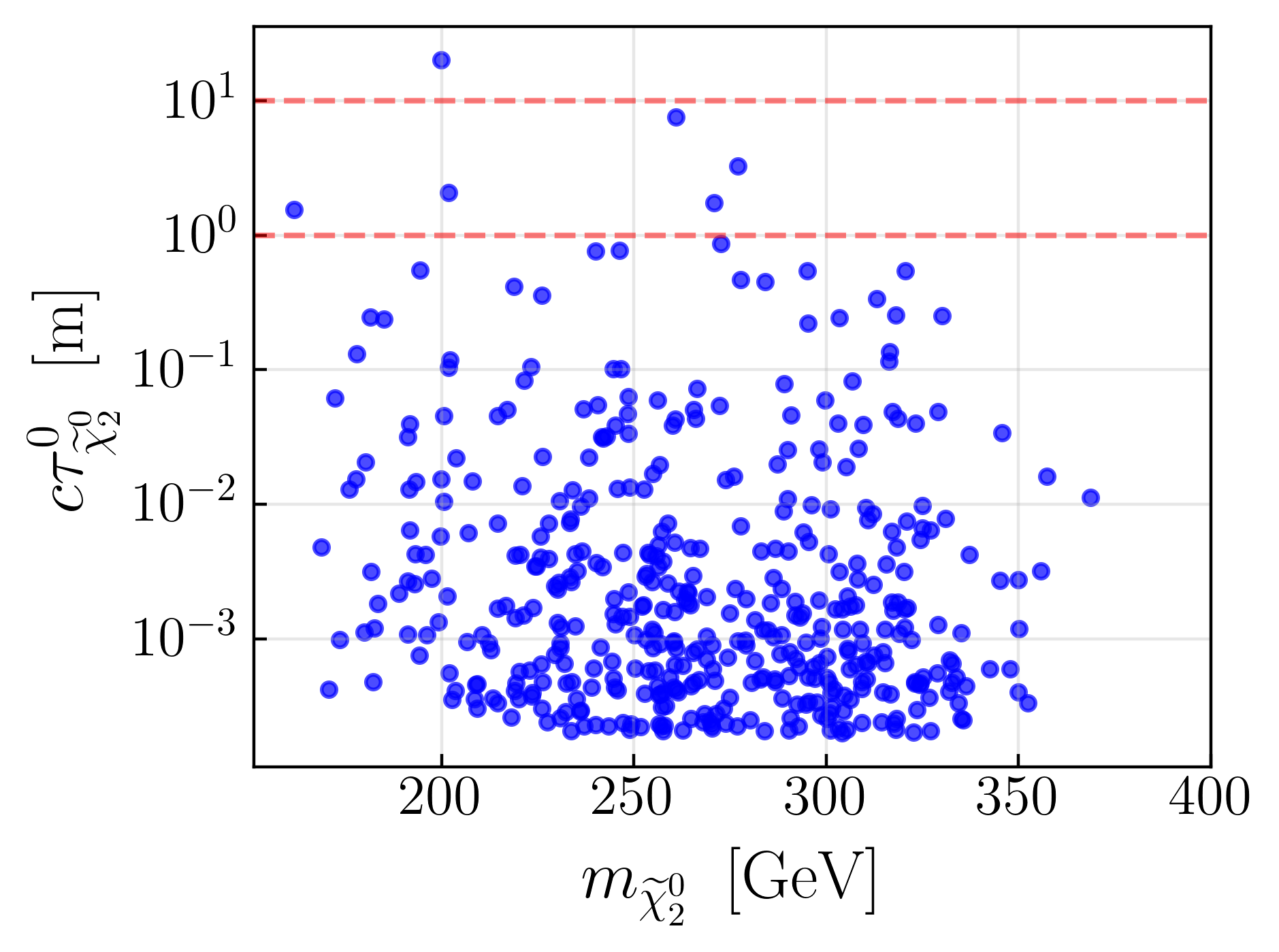}\hspace{1cm}\includegraphics[height=5.3cm,width=6.4cm]{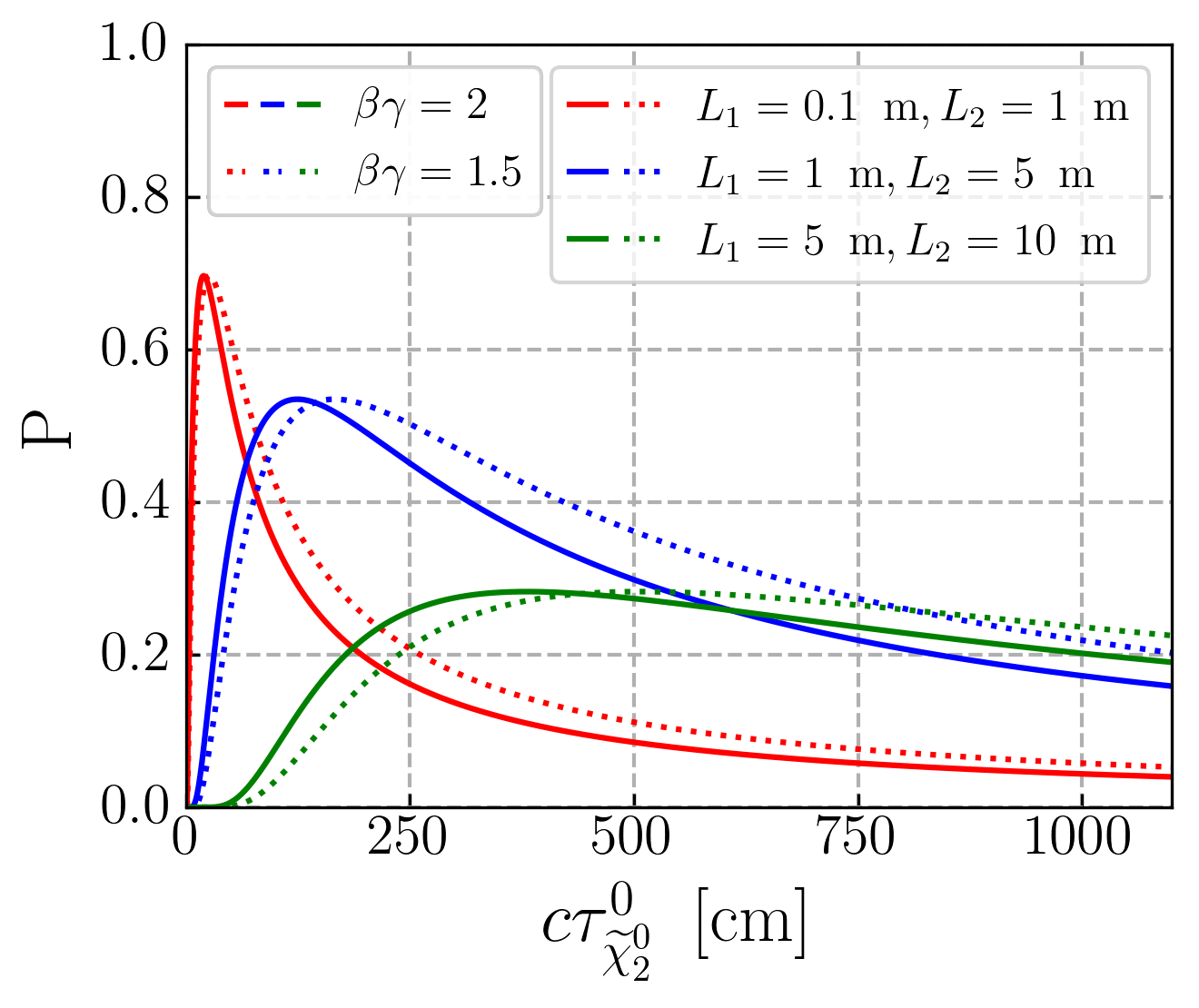}
	\caption{\textit{Left panel:} Decay length of the LLP as a function of its mass. The top and bottom red horizontal lines corresponds to a decay length of  $10 ~{\rm m}$ and, $1~\mathrm{m}$ respectively. This  shows that the SM decay products of LLP can reach ECAL and can also traverse a few meters in HCAL. For lengths $\geq \mathcal{O}(10)$ m they can even reach the muon-detectors.
	\textit{Right panel:} Acceptance probability $P$ vs. decay length of the LLP $\lsptwo$, $c\tau_{\lsptwo}^{0}$ for three choices of $\{L_{1},L_{2}\}$: $\{0.1~\mathrm{m}, 1~\mathrm{m}\}$, $\{1~\mathrm{m}, 5~\mathrm{m}\}$, and $\{5~\mathrm{m}, 10~\mathrm{m}\}$, assuming a hypothetical scenario with $\beta\gamma = 2$ and $1.5$. Here, $L_{1}$ and $L_{2}$ are the inner and outer radii of the LLP sensitive detector volume.}
	\label{fig:ctau_llp}
\end{figure}

In Fig.~\ref{fig:ctau_llp}~(left), we illustrate allowed parameter space with $\Gamma_{\lsptwo} \leq 10^{-13}~\mathrm{GeV}$~(Fig.~\ref{fig:DW_chi2}, pink points) in the plane of mean proper decay length, $c\tau_{\lsptwo}^{0}$ vs. mass of the LLP $\lsptwo$. Here, $\tau_{\lsptwo}$ represents the mean proper lifetime of $\lsptwo$ and $c$ is the speed of light. For convenience, we refer to the mean proper decay lifetime ($\tau^0$) as just ``lifetime" and the mean proper decay length ($c\tau^0$) ``decay length”, unless stated otherwise. The decay length for $\lsptwo$ is relatively large, $c\tau^0_{\lsptwo} \gtrsim 10~\mathrm{cm}$, for a considerable fraction of allowed parameter space. The decay length of $\lsptwo$ in the laboratory frame is given by
\begin{equation}
    l_{\lsptwo} = \beta \gamma c\tau_{\lsptwo}^{0}
\end{equation}
where, $\gamma = (1 - \beta^{2})^{-1/2} $ is the relativistic factor, $\beta = |\Vec{p}|/E = v/c$ is the boost, $v$ is the velocity, $E$ is the energy and $|\Vec{p}|$ is the momentum of the particle in the laboratory frame. The acceptance probability of the LLP $\lsptwo$ decaying within distance $L_1 < L < L_2$ inside the detector can be then expressed as,
\begin{equation}
\begin{split}
    &P = \int_{L_1}^{L_2} \frac{1}{l_{\lsptwo}}e^{-L/l_{\lsptwo}} \,dL \\
    &\Rightarrow  P (L_1,L2,\beta) \simeq \exp{\left({\frac{-L_1}{\beta\gamma c\tau_{\lsptwo}^{0}}}\right)} - \exp{\left({\frac{-L_2}{\beta \gamma c\tau_{\lsptwo}^{0}}}\right)},
    \label{eq:acceptance_prob}
\end{split}    
\end{equation}
where, $L_{1}$ and $L_{2}$ are the inner and outer radii of the detector volume that are sensitive to the LLP decay. Correspondingly, the number of observed LLP decays is given by,
\begin{equation}
    N_{LLP} = \mathcal{L}\times \sigma_{\text{signal}} \times Br \times  P(L_1,L2,\beta)\times \epsilon_{\text{reco}},
    \label{eq:num_llP_decay}
\end{equation}
where, $\mathcal{L}$ represents the integrated luminosity, $\sigma_{signal} \times Br$ represents the event production rate and $\epsilon_{\mathrm{reco}}$ corresponds to the signal efficiency. From Eq.~(\ref{eq:acceptance_prob}) it is clear that the acceptance probability decreases for relatively large decay lengths $c\tau_{\lsptwo}^{0} > L_{2}$. We illustrate the variation of acceptance probability with $c\tau_{\lsptwo}^{0}$ in Fig.~\ref{fig:ctau_llp}~(right), for three different choices of $\{L_{1},L_{2}\}$: $\{0.1~\mathrm{m}, 1~\mathrm{m}\}$, $\{1~\mathrm{m}, 5~\mathrm{m}\}$, and $\{5~\mathrm{m}, 10~\mathrm{m}\}$, assuming $\beta\gamma = 2$ and $1.5$. For $\{L_{1} = 0.1~\mathrm{m},L_{2}=1~\mathrm{m}\}$, the highest acceptance probability is observed for $c\tau_{\lsptwo}^{0} \sim 0.2~\mathrm{m}$. Similarly, detector volumes at larger radii illustrate maximal acceptance probability at larger decay lengths. 

The momentum resolution at the tracker is better than the energy resolution at the calorimeters for charged tracks~\cite{deFavereau:2013fsa}. Hence, the tracker enables more efficient identification of the charged tracks from LLP decays and reconstruction of the displaced secondary vertex~(DSV). Keeping this in mind, in the present work, we restrict our analysis in the tracker region using single/di-lepton triggers and missing energy. In both CMS and ATLAS detectors, the tracker region extends to a radius of $L_{2} \sim 1~\mathrm{m}$. Therefore, we consider only such signal benchmark points where the decay length of $\lsptwo$, $c\tau_{\lsptwo}^{0} \lesssim 1~\mathrm{m}$, such that the majority of $\lsptwo$ decays occur inside the tracker region~(c.f. Fig.~\ref{fig:ctau_llp}~(right)). Considering these observations, we identify $3$ benchmark points with $\Gamma_{\lsptwo} ({\rm in~GeV})\sim 10^{-14}$~(BP1), $\sim 10^{-15}$~(BP2) and $\sim 10^{-16}$~(BP3). In Table~\ref{Tab:BP}, we present the input parameters, along with masses, decay widths, and branching rates of the relevant electroweakinos and Higgs bosons, for BP1, BP2, and BP3.

\begin{figure}[!t]
    \centering
    \includegraphics[height=5.2cm,width=7cm]{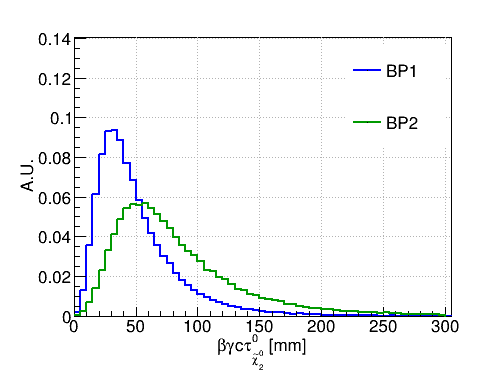}
    \caption{Distributions for decay length of the LLP $\lsptwo$ produced via $pp \to (\lspthree \to {\bf \lsptwo} h)\chonepm$ at $\sqrt{s}=14~\mathrm{TeV}$ for signal benchmark BP1$\{m_{\lspthree}=760~\mathrm{GeV}, m_{\lsptwo} = 217~\mathrm{GeV}, c\tau_{\lsptwo}^{0} = 17.5~\mathrm{mm}\}$ and BP2$\{m_{\lspthree}=791~\mathrm{GeV}, m_{\lsptwo} = 193~\mathrm{GeV}, c\tau_{\lsptwo}^{0} = 26~\mathrm{mm}\}$ are shown as blue and green solid lines, respectively.}
    \label{fig:decay_length_BP1_BP2}
\end{figure}

The decay of particles with a typical lifetime $\tau_{i}$ follows an exponential distribution,
\begin{equation}
    N(t) = N_{0} \exp{\left(-t/\tau_{i}\right)},
\end{equation}
where, $N(t)$ represents the number of undecayed particles after time $t$, and $N_{0}$ is the total number of particles produced with lifetime $\tau_{i}$. Accordingly, the $\lsptwo$'s in our signal benchmarks can undergo decays in different segments of the detector depending on the boost ($\beta$) and decay length ($\tau_{\lsptwo}^{0}$) where the latter is inversely correlated to $\Gamma_{\lsptwo}$. For illustration, we present the distributions for $l_{\lsptwo}$, where $\lsptwo$ is produced via $pp \to (\lspthree \to \lsptwo H_1)\chonepm$, for BP1~(blue solid) and BP2~(green solid), in Fig.~\ref{fig:decay_length_BP1_BP2}. These benchmarks have $c\tau_{\lsptwo}^{0}=~$17.5~mm~(BP1) and 26~mm~(BP2), respectively, thereby furnishing a relatively large acceptance probability in the tracker volume. We would like to note that particles with larger decay lengths can also undergo decay within the tracker region. However, the fraction of such decays would be small and warrants a separate study of its own~(c.f.~\cite{Banerjee:2019ktv}), which is beyond the scope of the present work.

As discussed previously, the decay of $\lsptwo \to \lspone b\bar{b}$ within the tracker region leads to charged tracks that originate from displaced secondary vertex~(DSV) corresponding to the LLP $\lsptwo$. An important parameter relevant for the reconstruction of DSV is the transverse impact parameter $|d_{0}|$,
\begin{equation}
|d_0| = \frac{\left|x_d^{track}p^{track}_y - y_d^{track}p^{track}_x \right|}{p^{track}_T},
\end{equation}
where, $\{x_{d}^{track},y_{d}^{track}\}$ are the track coordinates in the transverse plane passing through the primary interaction vertex~(PIV), $p_{x}^{track}$ and $p_{y}^{track}$ are the x- and y-components of the track momentum and $p_{T}^{track} = \sqrt{{p_{x}^{track}}^2 + {p_{y}^{track}}^{2}}$. In the present scenario, the displaced charged tracks from $\lsptwo \to \lspone b\bar{b}$ in BP1, BP2 and BP3 can feature a typically large $|d_{0}| \gtrsim \mathcal{O}(1)~\mathrm{cm}$, which is indicative of a DSV.

Note that the analysis strategy considered in this work requires separate tackling of the prompt and long-lived objects. Motivated from studies by the ATLAS and CMS collaboration in Refs.~\cite{CMS:2012axw,ATLAS:2012cdk}, we consider final state objects with transverse impact parameter $d_{0} \gtrsim 2~\mathrm{mm}$ to be long-lived while those with $d_{0} \lesssim 2~\mathrm{mm}$ as prompt.

\subsection{Benchmark points and analysis setup}

Having discussed the generic features of the LLP $\lsptwo$ in a collider environment, we will move on to study the projected capability of the HL-LHC to probe the NMSSM parameter space of interest through LLP searches in direct electroweakino production of Eq.~(\ref{eq:cascade}). As discussed previously, to this end, we perform a detailed collider study of three different benchmark points BP1, BP2, and BP3~(Table~\ref{Tab:BP}) chosen from the current allowed parameter space. We use \texttt{PYTHIA8}~\cite{Sjostrand:2006za,Sjostrand:2007gs} to simulate the signal process in Eq.~(\ref{eq:cascade}). The signal process is mediated through a promptly decaying $WZ/WH_{1}$ in addition to the late decaying $\lsptwo$'s, leading to a variety of prompt SM objects in the final state which could be potentially triggered upon. The list of dominant backgrounds would vary according to the trigger choice. The different viable triggers and the associated backgrounds will be discussed in details in Section~\ref{sec:signal_trigger}. We use the \texttt{Madgraph5-aMC@NLO-2.7.3}~\cite{Alwall:2014hca} framework to simulate the background events at the parton-level, with subsequent showering and hadronization being performed using \texttt{PYTHIA8}. The HL-LHC detector response is simulated using \texttt{Delphes-3.5.0}~\cite{deFavereau:2013fsa} using the default HL-LHC detector card~\cite{HL-LHC_card:Online}. 

As our analysis relies on tracks originating from the LLP, we do not cluster jets but rather use {\tt Delphes} collections both for generator and reconstructed level objects within our analysis. We separate out the prompt objects like leptons, which are primarily used for event selection. The main analysis deals with displaced `particle-flow' tracks. The generator level charged particles, estimated with a good resolution, have a finite probability to be reconstructed as tracks. We have checked that there is no overlap between reconstructed leptons and generator level charged particles with $|\eta| < 4 ~\&~ p_{T} >10 {\rm~GeV}$ in our analysis. At the stage of particle propagation, only smearing on the norm of the transverse momentum vector is applied, assuming a perfect angular resolution on tracks. In the mentioned updated Delphes module, a dedicated filter is used~\cite{Selvaggi2017delphesdelphesD} to enhance the tracking performance along with momentum resolution. This tackles inefficiencies in boosted, dense environments.

We would like to note that offline tracking efficiencies have not been included in our analysis due to the lack of definitive knowledge of the HL-LHC performance. Therefore, our results need to be considered as conservative estimates and can certainly be improved upon.

\begin{table*}[htb] 
\setlength{\tabcolsep}{12pt}
\centering
 \begin{tabular}{l c c c} 
 \hline
 & BP1 & BP2 & BP3 \\ [0.5ex]
 \hline
$\lambda$ & $5.15\times 10^{-3}$ & $5.85\times 10^{-3}$ & $1.67\times 10^{-4}$ \\
$\kappa$ &$6.12\times 10^{-4}$ & $5.854\times 10^{-4}$ & $2.07 \times 10^{-5}$\\
$A_\lambda$ [GeV] & 5279 & 2110 & 9705 \\
$A_\kappa$ [GeV] &-32   & -510 & -21 \\
$ \mu$  [GeV] & 743.05 &  775.05 & 688.05 \\
$\text{tan}\beta$ & 25.098 & 36.32 & 44.67\\
$M_1$ [GeV]& 218.39 & 194.4  & 238.8\\
$M_2$ [GeV]& 3909 & 3709 & 2789\\
$M_3$ [GeV]& 4219 & 4371 & 5465\\
\hline
$m_{\lspone}$ [GeV]& 180.17& 158.08 & 173.76\\
$m_{\lsptwo}$ [GeV] & 216.76& 193.00 & 236.93\\
$m_{\lspthree}$ [GeV] & 759.62& 790.67 & 703.55\\
$m_{\lspfour}$ [GeV] & 760.42& 791.80 & 704.94\\
$\mchonepm$ [GeV] & 758.43& 789.72 & 702.37\\
$m_{H_1}$ [GeV] & 126.31& 122.52 & 124.54\\
$m_{H_2}$ [GeV] & 168.43& 143.11 & 164.7\\
$m_{A_1}$ [GeV] & 92.0& 108.90 & 73.19\\
\hline
$\Gamma_{\lsptwo}$ [GeV] &  $1.11\times10^{-14}$ & $7.69\times10^{-15}$ & $3.85\times 10^{-16}$\\
$\Gamma_{\lspthree}$ [GeV] & 0.4847 &  0.5002 & 0.4367\\
$\Gamma_{\lspfour}$ [GeV] & 0.4571 &  0.4755 & 0.4088\\
\hline
$\sigma_{\rm NLO}$ [pb] & 1.56 & 1.15 & 2.15\\ 
\hline
BR($\lsptwo \rightarrow \lspone b \bar{b}$) & 0.528 & 0.63 & 0.34\\
BR($\lsptwo \rightarrow \lspone j \bar{j}$) & 0.1834 & 0.074  & 0.3602 \\
BR($\lsptwo \rightarrow \lspone \tau^+  \tau^-$) & 0.12 & 0.177 & 0.0969\\
BR($\lsptwo \rightarrow \lspone \l^+  \l^-$) & 0.085 & 0.014 &  0.176\\
BR($\lspthree \rightarrow \lsptwo H_1 $) & 0.79 & 0.704 &  0.816\\
BR($\lspthree \rightarrow \lsptwo Z$) & 0.204 & 0.24 & 0.184\\
BR($\lspfour \rightarrow \lsptwo Z$) & 0.7834 & 0.74 & 0.811\\
BR($\lspfour \rightarrow \lsptwo H_1$) & 0.215 & 0.24 & 0.189\\
BR($\chonemp \rightarrow \lsptwo W$) & 0.994 & 0.995 & 0.999\\
\hline
\hline
\end{tabular}
\caption{The input parameters, Higgs boson and electroweakino mass spectrum, branching ratios of electroweakinos, decay width and decay length of the LLP $\lsptwo$, and the production cross-section for the process $pp \to \lspthree \chonepm + \lspfour \chonepm$ at $\sqrt{s}=14~{\rm TeV}$, for BP1, BP2, and BP3.}
\label{Tab:BP}
\end{table*}

\subsection{Signal trigger}
\label{sec:signal_trigger}
\begin{figure}[!t]	
	\centering
	\includegraphics[height=5.2cm,width=7cm]{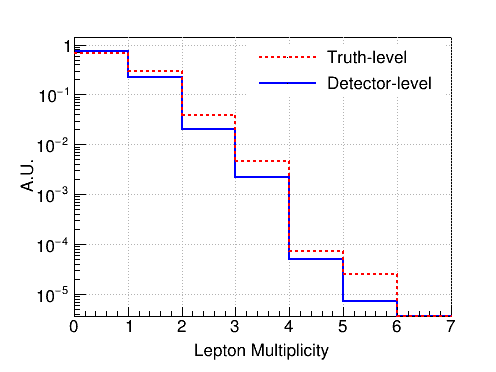}
	\caption{Distribution for lepton multiplicity $n_{\ell}$~($\ell=e,\mu$) from promptly decaying $WZ/WH_{1}$ pair produced in the process $pp \to (\chonepm \to W^{\pm}\lsptwo)(\lspthree/\lspfour \to Z/H_{1} \lsptwo)$ at the HL-LHC. The truth level and detector-level distributions are shown in red solid and blue solid, respectively.}   
	\label{fig:lepton_multiplicity}
\end{figure}

Before moving on to discuss the strategy to reconstruct the DSV's associated with $\lsptwo$, let us examine the prompt components in the signal process. The $WZ/WH_{1}$ pair produced through $pp \to (\chonepm \to W^{\pm}\lsptwo)(\lspthree/\lspfour \to Z/H_{1} \lsptwo)$ decay promptly and can lead to various different SM final states which could be triggered upon. In Fig.~\ref{fig:lepton_multiplicity}, we present the multiplicity $n_{\ell}$ of isolated prompt leptons $\ell$~($= e$, $\mu$) that can originate from the decay of $WZ/WH_{1}$ pair in the signal process in Eq.~(\ref{eq:cascade}) at truth level (red, dotted line) and detector level (blue solid line). At the detector-level, an isolated lepton is required to satisfy,
\begin{equation}
   \frac{\sum\limits_{i \neq l} {p_{\text{T}}(i)}^{\Delta R<0.3}} {p_{T,l}} < 0.1,\quad l = e, \mu
\end{equation}
where, $\sum\limits_{i \neq l} {p_{\text{T}}(i)}^{\Delta R<0.3}$ represents the sum of transverse momenta of all objects (excluding the lepton candidate) with $p_{T} > 2~\mathrm{GeV}$ within a cone of radius $\Delta R < 0.3$ centred around the candidate lepton, $\Delta R = \sqrt{\Delta\eta^{2} + \Delta \phi^{2}}$ where $\Delta \eta$ and $\Delta \phi$ are the pseudorapidity and azimuthal angle differences and $p_{T,l}$ is the transverse momentum of the candidate lepton. We would like to note that these isolated leptons are also required to satisfy $d_{0} < 2~\mathrm{mm}$.

The single prompt lepton final state in Fig.~\ref{fig:lepton_multiplicity} can originate when the $W$ decays leptonically, $W\to \ell^{\prime}\nu$~($\ell^{\prime} = e, \mu, \tau$) while the $Z/H_{1}$ decays hadronically. Similarly, the final state with two prompt leptons can arise when $W$ decays hadronically~($W \to jj$) while the $Z/H_{1}$ decays via leptonic modes. The scenario with $n_{\ell}=3$ can arise when (a) $W\to \ell^{\prime}\nu, Z \to \ell^{\prime} \ell^{\prime}$ (b) $W\to \ell^{\prime}\nu, H_{1} \to \ell^{\prime} \ell^{\prime}$ (c) $W\to \ell^{\prime}\nu, H_{1} \to (W\to\ell^{\prime}\nu)(W^{*} \to \ell^{\prime}\nu)$ (d) $W\to \ell^{\prime}\nu, H_{1} \to (Z \to \ell^{\prime} \ell^{\prime})(Z^{*} \to jj)$ etc. Accordingly, we consider two different analysis categories corresponding to different signal triggers, $n_{\ell} = 1$ and $n_{\ell} = 2$. The $n_{\ell} =3$ signal category is ignored due to smaller production rates relative to the other two. For $n_{\ell} = 1$, we require the isolated prompt lepton to satisfy $p_{T,\ell} > 30~\mathrm{GeV}$. Recall that the signal final state contains two long-lived $\lsptwo$ each decaying into $\lsptwo \to \lspone + b\bar{b}$. Therefore, in addition to prompt leptons, we will have $b~\mathrm{jets} + \met$.
The dominant backgrounds in this signal category are semileptonic $t\bar{t}$ and $W+\mathrm{jets}$.  In the $n_{\ell}=2$ signal category, we impose $p_{T,\ell_{1}} > 30~\mathrm{GeV}$ and $p_{T,\ell_{2}} > 20~\mathrm{GeV}$ where $p_{T,\ell_{1}} > p_{T,\ell_{2}}$. The dominant backgrounds are dileptonic $t\bar{t}$ and $2\ell+jets$, where $jets$ mainly include $b$ and $c$ jets. 
We also require the isolated leptons to lie within $|\eta| < 4.0$ and impose a lower threshold on the missing transverse energy $\met > 50~\mathrm{GeV}$ at the event selection stage. Alternatively, jet triggers can be used instead of lepton triggers since the $WZ/WH_{1}$ pair in the signal can predominantly decay via hadronic modes. Choosing an optimized event triggering criteria for the online trigger systems $viz$ the Level-1 (L1) trigger and the HLT is among the most critical steps in any analysis (c.f. Refs.~\cite{Bhattacherjee:2020nno, Bhattacherjee:2021qaa} and references therein). The choice of efficient triggers is more pertinent for the L1 to ensure that the events of interest ($viz$ the LLPs in the present analysis) are not lost forever. The event selection rates at the HLT are, on average, an order of magnitude smaller than at the L1 system. Therefore, typically stronger thresholds are applied to the HLT system to ensure consistent event rates. In this regard, triggering on leptons is advantageous due to similar thresholds at the L1 trigger and the HLT, inclusiveness, and less susceptibility to pile-up effects. For the case of a single isolated muon (electron), the values of L1 trigger seed of $p_{T} > 22~\mathrm{GeV}$ ($28~\mathrm{GeV}$) and $|\eta| < 2.4$, are pretty similar to the threshold at the HLT viz. $p_{T} > 24~\mathrm{GeV}$ ($32~\mathrm{GeV}$)~\cite{CERN-LHCC-2020-004, Collaboration:2759072}.
On the other hand, HLT thresholds on $p_{T}$ and the sum of transverse momenta $H_{T}$ for the jet(s) are harder compared to their L1 counterparts. Therefore, jet-triggered events are vulnerable to considerable efficiency loss at the HLT. Furthermore, the jet thresholds are also strongly sensitive to the level of pile-up and the high pile-up environment at the HL-LHC can degrade jet energy resolution leading to depleted trigger efficiencies~\cite{CERN-LHCC-2020-004}. Optimal jet trigger rates require the implementation of dedicated pile-up mitigation techniques, which is beyond the scope of our work. Therefore, for simplicity, we adhere to lepton triggers only.
The signal triggers and the corresponding selection cuts are summarized in Table~\ref{tab:signal_trigger}.

\begin{figure}[!t]
    \centering
    \includegraphics[height=5.2cm,width=7cm]{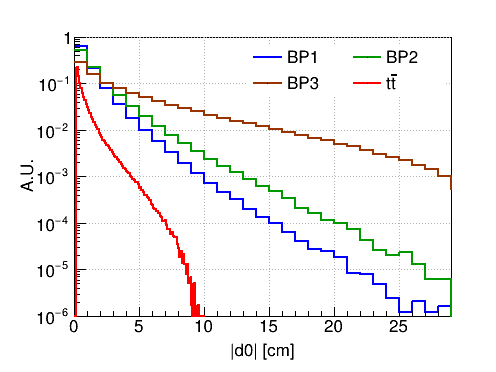}\hspace{1 cm}
    \includegraphics[height=5.2cm,width=7cm]{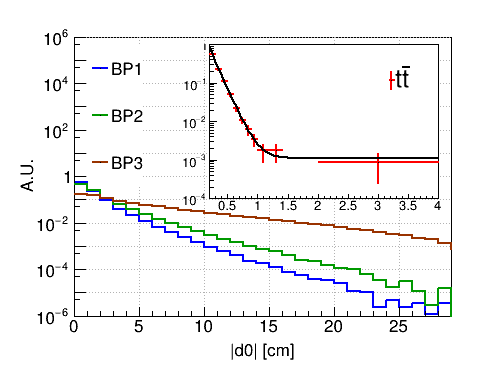}
    \caption{Distribution of transverse impact parameter $|d_{0}|$ for all tracks with $p_{T} > 1~\mathrm{GeV}$ and $|\eta| < 4$ in the $pp \to \chonepm\lspthree/\lspfour \to (\chonepm \to \lsptwo + W^{\pm}, \lsptwo \to \lspone + b\bar{b})(\lspthree/\lspfour \to \lsptwo + Z/H_{1}, \lsptwo \to \lspone + b\bar{b})$ channel corresponding to BP1~(blue), BP2~(green) and BP3~(brown), at the HL-LHC. \textit{Left panel:} Events pass the trigger choice $n_{\ell} = 1$. The corresponding distributions for the semileptonic $t\bar{t}$ background is shown in red color. \textit{Right panel:} Events pass the trigger choice $n_{\ell} = 1$ and have at least one displaced secondary vertex. The distributions for the semileptonic $t\bar{t}$ background is shown in red color in the figure inset.}
    \label{fig:d0_all_track}
\end{figure}

\begin{table}[htb]
    \centering
    \begin{tabular}{|c|c|}\hline
    \multicolumn{2}{|c|}{Signal triggers} \\ \hline\hline
    \multicolumn{2}{|c|}{$|d_{0}| < 2~\mathrm{mm}$} \\ \hline
    $n_{\ell} = 1$ & $n_{\ell} = 2$ \\ \hline
        $p_{T,\ell_{1}} > 30~\mathrm{GeV}$ & $p_{T,\ell_{1}} > 30~\mathrm{GeV}$ \\ 
         & $p_{T,\ell_{2}} > 20~\mathrm{GeV}$  \\  \hline
         \multicolumn{2}{|c|}{$\met > 50~\mathrm{GeV}$} \\ \hline
    \end{tabular}
    \caption{Summary of signal triggers and the basic selection cuts. These triggers are only applied to prompt objects. Tracks with $|d_{0}| < 2~\mathrm{mm}$ are classified within the prompt category.}
    \label{tab:signal_trigger}
\end{table}

\subsection{Reconstructing the displaced secondary vertex from LLP $\lsptwo$}
As discussed earlier, the cascade decay channel for the directly produced $\lspthree\chonepm/\lspfour\chonepm$ pair results into two $\lsptwo$'s in addition to other prompt SM candidates. The two LLP candidates can in principle lead to two displaced secondary vertices. In the signal, the tracks with larger transverse impact parameters are expected to originate from these two displaced secondary vertices. To reconstruct the final DSVs, we retrace the tracks with $d_{0} \geq 2~\mathrm{mm}$. In Fig.~\ref{fig:d0_all_track}~(left), we present the distributions for $d_{0}$ for the signal benchmarks BP1, BP2, and BP3. Here, we include all tracks with $p_{T} > 1~\mathrm{GeV}$ and $|\eta| < 4.0$ in events that pass the selection cuts corresponding to the signal trigger $n_{\ell}=1$. The corresponding distributions for the dominant semileptonic $t\bar{t}$ background are also illustrated in the same figure. We observe from Fig.~\ref{fig:d0_all_track}~(left) that the tail of the distributions for the signal process shifts to larger $|d_{0}|$ values with decreasing $\Gamma_{\lsptwo}$. For BP3 the fraction of events~($n_{\rm frac}$) above $|d_{0}| > 10~\mathrm{cm}$, $n_{\rm frac} \sim 0.02$. In this range of $|d_{0}|$ the fraction is considerably low for BP1~($n_{\rm frac} \sim 0.001$) and BP2~($n_{\rm frac} \sim 0.005$). Interestingly, the $|d_{0}|$ distributions for the semileptonic $t\bar{t}$ background extends all the way up to $|d_{0}| \sim 10~\mathrm{cm}$. This happens due to long-lived mesons like $K_{s}^{0}$, $\Lambda$, $D$ etc. produced from $b$ hadrons. Therefore, it is essential to explore other features of the LLP-specific topology which can reduce the backgrounds. One such entity that is largely exclusive to the phenomenology of long-lived decays is the displaced secondary vertex. As such, our next objective is to reconstruct the secondary vertices associated with the LLP $\lsptwo$. We will also explore various observables that are contingent on the reconstructed DSVs, optimizing the selection cuts on them, and revisit the $|d_{0}|$ distributions afterward.

In an ideal scenario, tracks that arise from the same secondary vertex are expected to share a common point of origin $\{x_{0},y_{0},z_{0}\}$. Correspondingly, we allocate tracks with $d_{0} \geq 2~\mathrm{mm}$ whose point of origin are within $\{|\Delta x| < 1~\mathrm{mm}$, $|\Delta y| < 1~\mathrm{mm}$, $|\Delta z| < 1~\mathrm{mm}\}$ of each other, to a reconstructed vertex. Among them, the ones that contain at least 3 tracks are classified as a DSV. Having reconstructed the DSVs, let us revisit the distributions for $|d_{0}|$. We redraw the distributions for $|d_{0}|$ in Fig.~\ref{fig:d0_all_track}~(right), similar to that in Fig.~\ref{fig:d0_all_track}~(left), except now with only those events which have at least one reconstructed DSV. 

Let us also note the following important fact about the $t\bar{t}$ background. Imposing the requirement for a reconstructed DSV leads to a major depletion in the $d_{0}$ distributions for the semileptonic $t\bar{t}$ background. It falls sharply before it reaches $|d_{0}| \sim 2~\mathrm{cm}$ and suffers from substantial statistical uncertainty in the tail. Therefore, we extrapolate the shape of $|d_{0}|$ for semileptonic $t\bar{t}$ background using $5\rm{M} ~t\bar{t}$ events in Fig.~\ref{fig:d0_all_track}~(right) and is shown by the solid black line. To ensure consistency, we generate additional $6.5$~M $~t\bar{t}$ events, and the extrapolated function derived in the previous step matches with the $|d_{0}|$ distributions drawn for this new sample. Note that the long tail for the $|d_{0}|$ distribution in case of the $t\bar{t}$ background, is an artifact of rarity of events with large decay length in SM.

\subsection{LLP-specific observables at the detector level}
With an enhanced tracking algorithm, ATLAS shows a good reconstruction efficiency even for displaced tracks produced at a large radius within 30 cm from the primary interaction vertex \cite{ATLAS:2017kyn}. In order to reconstruct a displaced vertex, first, its tracks from that vertex need to be successfully reconstructed. Tracks originating far from the center of the detector tend to have higher values of $d_{0}$. Standard track reconstruction have low efficiency for large $d_{0}$ values. In the following, we construct a few variables without solely relying on $d_{0}$ to eliminate the background.

\begin{figure*}[!t]
    \centering
    \includegraphics[height=5.2cm,width=7cm]{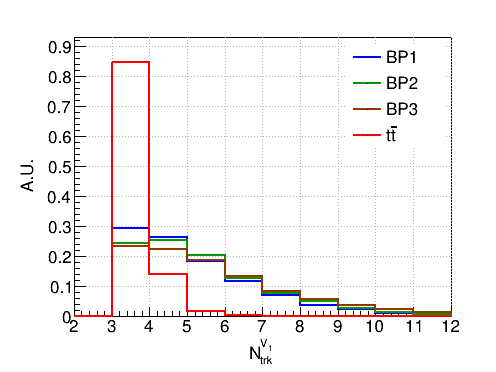}\hspace{1 cm}
    \includegraphics[height=5.2cm,width=7cm]{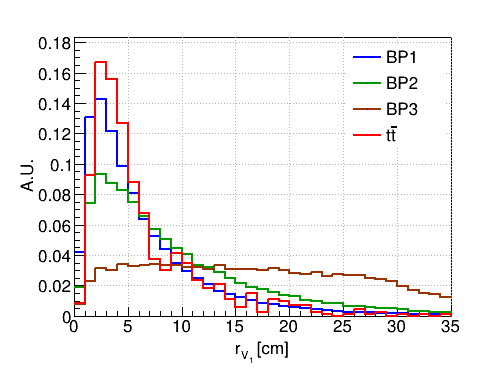}\\
    \includegraphics[height=5.2cm,width=7cm]{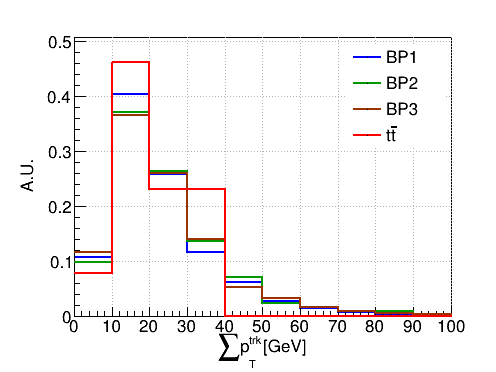}
    \caption{Distributions for track multiplicity of DSV $V_{1}$, $N^{V_{1}}_{\rm trk}$~(upper-left panel), radial distance of $V_{1}$ from the primary interaction vertex, $r_{V_{1}}$~(upper-right panel), and sum of transverse momentum of all tracks in $V_{1}$, $\sum{\rm {p_T^{trk}}}$~(lower panel), in the $pp \to \chonepm\lspthree/\lspfour \to (\chonepm \to \lsptwo + W^{\pm}, \lsptwo \to \lspone + b\bar{b})(\lspthree/\lspfour \to \lsptwo + Z/H_{1}, \lsptwo \to \lspone + b\bar{b})$ channel corresponding to BP1~(blue), BP2~(green) and BP3~(brown) at the HL-LHC. Distributions for the semileptonic $t\bar{t}$ background are shown in red.}
    \label{fig:DSV_feature_1}
\end{figure*}

We refer to DSVs with the highest and $2^{nd}$ highest track multiplicity as $V_{1}$ and $V_{2}$, respectively. For illustration, we show the track multiplicity of $V_{1}$, referred to as $N^{V_{1}}_{\rm trk}$, for BP1, BP2, and BP3, in Fig.~\ref{fig:DSV_feature_1}~(upper-left). These distributions are presented for the $n_{\ell} = 1$ signal trigger region, summarized in Table~\ref{tab:signal_trigger}.  We observe that $N^{V_{1}}_{\rm trk}$ can reach up to $\sim 5-6$ for a considerable fraction of signal events in all three benchmark points. On the other hand, $N^{V_{1}}_{\rm trk}$ reaches only up to $\sim 3$ in the semileptonic $t\bar{t}$ process, which is the dominant background when the signal trigger is $n_{\ell} = 1$. Accordingly, we optimize $N^{V_{1}}_{\rm trk}$ to improve signal-to-background discrimination.

Another parameter of interest is $ r_{V_{1}}$ which represents the radial distance of $V_{1}$ from the PIV. $r_{V_{1}}$ is computed as $r_{V_{1}} = \sqrt{X_{V_{1}}^2 + Y_{V_{1}}^2 + Z_{V_{1}}^2}$, where $\{X_{V_{1}},Y_{V_{1}},Z_{V_{1}}\}$ are the coordinates of the reconstructed DSV $V_{1}$ in a reference frame centred at PIV = $\{0,0,0\}$. In Fig.~\ref{fig:DSV_feature_1}~(center), we illustrate $r_{V_{1}}$ for the signal benchmarks and semileptonic $t\bar{t}$ background, considering the $n_{\ell} = 1$ signal trigger. The radial distance of the DSV from PIV is inversely proportional to the decay width of LLP in addition to the effect of Lorentz factor $\beta\gamma$. This behaviour is illustrated in Fig.~\ref{fig:DSV_feature_1}~(upper-right) where the distributions for $ r_{V_{1}}$ get flatter and the tail shifts to larger values as the decay length of $\lsptwo$ is increased. In the case of BP1, where $\Gamma_{\lsptwo} \sim 10^{-14}~\mathrm{GeV}$, $ r_{V_{1}}$ peaks roughly at $2~\mathrm{cm}$. As we move to BP2 where $\Gamma_{\lsptwo}$ is smaller by an order of  magnitude, the peak position shifts noticeably, however the overall distributions get flatter. At further lower values of $\Gamma_{\lsptwo} \sim 10^{-16}~\mathrm{GeV}$ corresponding to BP3, we observe a considerable alteration in the distribution. The corresponding distribution for the semileptonic $t\bar{t}$ background peaks at a much lower value $r_{V_{1}} \sim 3~\mathrm{cm}$. Overall, this observable demonstrates potential not only as a background discriminator but also as an excellent identifier of variations in the decay width of the LLP. Consequently, we optimize the selection cuts on $r_{V_{1}}$ such that the signal significance $S/\sqrt{B}$ is maximized, where $S$ and $B$ are the signal and background yields at the HL-LHC. In addition to $N^{V_{1}}_{\rm trk}$ and $r_{V_{1}}$, we also optimize the selection cut on the sum of transverse momentum of all tracks associated with $V_{1}$, represented as $\sum{\rm {p_T^{trk}}}$. We present the distributions for $\sum{\rm {p_T^{trk}}}$ in Fig.~\ref{fig:DSV_feature_1}~(bottom). The $\sum{\rm {p_T^{trk}}}$ distributions for both signal and the $t\bar{t}$ background peaks in the same region of $15~{\rm GeV}$, however the background falls sharply compared to the signal. While the backgrounds become negligible at $\sum{\rm {p_T^{trk}}} \gtrsim 40~\mathrm{GeV}$, the signal tail extends far beyond. Correspondingly, we optimize the upper limit on $\sum{\rm {p_T^{trk}}}$ in our cut-based analysis. 

\subsection{Optimized signal regions and cut flow}

\begin{table}[!t]
\centering\scalebox{0.73}{
\begin{tabular}{|c|c||c|c|c||c|c|c|c|c||c|c||c|c|c|c|} 
\cline{1-14}
\multicolumn{2}{|c|}{} & \multicolumn{3}{c||}{Prompt sector} & \multicolumn{5}{c||}{SR1} &  \multicolumn{2}{c||}{SR2} &  \multicolumn{2}{c||}{SR3}   \\
\multicolumn{2}{|c|}{\multirow{3}{*}{}} & \multicolumn{3}{c||}{${\rm |d_{0}| < 2~mm }$} & \multicolumn{5}{c||}{${\rm |d_{0}| > 2~mm }$} &  \multicolumn{2}{c||}{${\rm |d_{0}| > 4~mm }$} &  \multicolumn{2}{c||}{${\rm |d_{0}| > 8~mm }$}   \\\cline{3-14} \cline{3-14}
\multicolumn{2}{|c|}{}& \makecell{${\rm {D_0^{\ell},D_Z^{\ell}}}$ \\ ${\rm < 2~ mm}$} & \makecell{${\rm {p_T^{\ell_{1,2}}}} >$ \\ ${\rm 30,20~GeV}$} &  \makecell{$\met$ \\ ${\rm > 50 ~GeV}$} & \makecell{${\rm {N_{trk}^{V1}}}$ \\ ${\rm > 3}$ }& \makecell{${\rm {N_{trk}^{V1}}}$ \\ ${\rm > 5}$ } & \makecell{$\sum{\rm {p_T^{trk}}}$ \\ ${\rm < 30 ~ GeV}$} & \makecell{${\rm {r_{V_1}}}$ \\ ${\rm > 4~ cm}$} & \makecell{${\rm {N_{trk}^{V2}}}$ \\ ${\rm > 5}$ } &  \makecell{${\rm {N_{trk}^{V1}}}$ \\ ${\rm > 3}$ } & \makecell{${\rm {N_{trk}^{V1}}}$ \\ ${\rm > 5}$ } &  \makecell{${\rm {N_{trk}^{V1}}}$ \\ ${\rm > 3}$ } & \makecell{${\rm {N_{trk}^{V1}}}$ \\ ${\rm > 5}$ } \\\hline
\multirow{4}{*}{1$\ell$} & BP1 & 563 & 421 & 409 & 300 & 134  & 102  & 53 & 8 & 242 & 96 & 164 & 56\\
                       & BP2 & 562 & 427 & 416 & 348 & 175  & 127  & 87 & 15 & 305 & 137 & 238 & 92  \\
                       & BP3 & 318 & 236 & 228 & 117 & 63 & 46  & 42 & 7 & 113 & 59 & 105 & 52 \\
         & ${\rm t_h t_\ell}$ & $2\times10^8 $  &$1.31\times10^8 $ &$1\times 10^8$ & 99936 & 1914 & 1367 & 684 & 0 
         & 2187 & 0 & 137 & 0 \\
\hline  \hline

\multirow{4}{*}{2$\ell$} & BP1 & 82  & 53  & 52  & 38  & 17 & 13 & 7 & 1 & 31 & 12 & 21 & 7 \\
                       & BP2 & 78  & 52  & 51  & 43  &  22 & 16  & 11 & 2 & 38 & 17 & 30 & 11 \\
                       & BP3 & 45  & 29  & 28  & 15  & 8 & 6  & 5 & 1 & 14 & 7 & 13 & 7\\
          & ${\rm t_\ell t_\ell}$ & $6.02 \times 10^{7}$ & $4.5 \times 10^{7}$ & $3.5 \times 10^{7}$ & 35712 & 624 & 406 & 281 & 0  & 812  & 0 & 62 & 0\\
\hline        
\end{tabular}
}
\caption{Selection cuts on lepton $p_{T}$~(i=1,2) and $\met$ for prompt candidates $|d_{0}| < 2~\mathrm{mm}$, and on $|d_{0}|$, $N_{\rm trk}^{V_1}$, $\sum{\rm {p_T^{trk}}}$, $r_{V_{1}}$ and $N_{\rm trk}^{V_{2}}$, for displaced objects $|d_{0}|>2\mathrm{~mm}$ are applied successively. Three Signal Regions have been specified based on the $|d_{0}|$ limits of 2, 4, 8 mm viz SR1, SR2, SR3. Selection cuts tabulated under prompt sector in each signal category are common to all corresponding signal regions. From the background rates, the displaced object $N_{\rm trk}^{V_1}$ is sufficient to optimize SR2 \& SR3. Signal and background rates are presented for $\sqrt{s}=14~\mathrm{TeV}$ LHC assuming $\mathcal{L}=3000~\mathrm{fb^{-1}}$.} 
\label{tab:cut_flow}
\end{table}

We present the optimized selection cuts on $p_{T,\ell_{1,(2)}}$ and $\met$ for the $n_{\ell}=1~(2)$ signal category, along with the corresponding signal and background rates for BP1, BP2, BP3 and the $t\bar{t}$ semileptonic ($t_{h}t_{\ell}$) and fully leptonic~($t_{\ell}t_{\ell}$) backgrounds at the HL-LHC in Table~\ref{tab:cut_flow}. These cuts are applied on prompt objects $|d_{0}| < 2~\mathrm{mm}$. Selection cuts on $|d_{0}|$, $N_{\rm trk}^{V_1}$, $\sum{\rm {p_T^{trk}}}$, $r_{V_{1}}$ and $N_{\rm trk}^{V_{2}}$ (Number of tracks from vertex $V_2$) are optimized for the long-lived objects in SR1. The $S/B$ ratio, where $S$ and $B$ are the signal and background yields at the HL-LHC, is $\sim \mathcal{O}(10^{-6})$ after the application of prompt category cuts for all signal benchmarks and signal categories. Imposing the requirement for at least one DSV with $N_{\rm trk}^{V_{1}} \geq 3$ using long-lived objects $|d_{0}| > 2~\mathrm{mm}$ improves $S/B$ to $\sim \mathcal{O}(10^{-3})$. Furthermore, this requirement leads to negligible event rates for $W+\mathrm{jets}$ and $Z+\mathrm{jets}$ in the $n_{\ell}=1$ and $n_{\ell}=2$ signal regions respectively, and can be safely ignored. The successive imposition of $N_{\rm trk}^{V_1} \geq 5$ in SR1 reduces the background rates further by a factor of $\sim 50$ while the signal rates for the three benchmarks reduces only by a factor of $\sim 2$. We eventually arrive at a negligible background scenario in SR1 on imposing $\sum p_{T}^{\rm trk} < 30~\mathrm{GeV}$, $r_{V_{1}} > 4~\mathrm{cm}$ and $N_{\rm trk}^{V_2}\geq$  5. We analyze two additional cases where the long-lived objects are required to satisfy a more stringent $|d_{0}|$ criteria $viz$ $|d_{0}| \geq 4~\mathrm{mm}$~(SR2) and $|d_{0}| \geq 8~\mathrm{mm}$~(SR3). In both cases, we observe that backgrounds can be suppressed with $N_{\rm trk}^{V_1} \geq 5$. Thus, set of cuts $\{$SR1, SR2, SR3$\}$ can effectively remove the backgrounds facilitating the exclusion/observation of BP1/BP2/BP3 at the HL-LHC in the $n_{\ell}=1$ and $2$ signal category, respectively.

\section{Outlook and conclusion}
\label{sec:conclusion}
In this work, we focus on the case of singlino-like light neutralino DM in the NMSSM framework. Implications from current collider and astrophysical constraints have been analyzed, and the allowed parameter space has been scrutinized in light of projected sensitivities in the future direct detection experiments. We consider an electroweakino mass spectrum where $\lsptwo$ has a dominant bino admixture, $\lspthree, \lspfour, \chonepm$ have a dominant higgsino composition, and $\lspfive,\chtwopm$ are wino-like.
In the allowed region of parameter space, there exist long-lived bino-like NLSP $\lsptwo$. The small decay width of this $\lsptwo$ being caused for, $\Delta M = m_{\lsptwo} - m_{\lspone} < m_{Z}$ which allows only 3 body decay for $\lsptwo$.   
Within the scope of the allowed parameter space of interest, the long-lived $\lsptwo$ can decay through $\lsptwo \to \lspone b\bar{b}$, $\lsptwo \to \lspone \tau^{+}\tau^{-}$, $\lsptwo \to \lspone j j$ or $\lsptwo \to \lspone \gamma$. The $\lsptwo$'s can appear in direct electroweakino searches at the LHC via cascade decays of heavier electroweakinos, and lead to displaced secondary vertices, which can be reconstructed in the tracker region of the LHC. In this work, we study the projected sensitivity for direct electroweakino production $pp \to \lspthree/\lspfour\chonepm \to (\lspthree/\lspfour \to Z/H_{1}\lsptwo)(\chonepm \to W^{\pm}\lsptwo)$ with $\lsptwo \to \lspone b\bar{b}$ at the HL-LHC. We choose three different signal benchmarks BP1, BP2, and BP3, from the currently allowed parameter space that features a long-lived $\lsptwo$. We perform a detailed collider analysis using the cut-and-count methodology while including signal and relevant backgrounds at the detector level. We note that the other decay modes for $\lsptwo$ $viz$ $\lsptwo \to \lspone \tau^{+}\tau^{-}/jj$ will also lead to charged tracks, which can be used to reconstruct the secondary vertices. We postpone the study of such final states to future analysis.

We consider two different signal categories, $n_{\ell}=1,2$ as discussed in Sec.~\ref{sec:Result}. To separate the signal from the background effectively, we use selection cuts on separate sets of observables in case of prompt and long-lived objects. Objects with transverse impact parameter $|d_{0}| < 2~\mathrm{mm}$ are classified as prompt, while those with $|d_{0}| \geq 2~\mathrm{mm}$ are categorized as long-lived. Prompt objects are used to trigger the events, while the displaced objects play the major role in discriminating against the backgrounds. We identify the signal regions SR1 with optimized selection cuts on $N_{\rm trk}^{V_1}$, the track multiplicity of $V_{1}$, $\sum p_{T}^{\rm trk}$, sum of transverse momentum of all tracks associated with $V_{1}$, $ r_{V_1}$, radial distance between $V_{1}$ and PIV, and $N_{\rm trk}^{V_2}$, track multiplicity for the second DSV. SR2 and SR3 are defined by optimizing the cuts on $|d_{0}|$, the minimum transverse impact parameter, and $N_{\rm trk}^{V_1}$. We show that with the choice of the three signal regions SR1, SR2, ans SR3, one can completely suppress the background, and the signal benchmarks BP1, BP2, and BP3, can be probed at the HL-LHC. Similar analysis can be extended to other points in the allowed parameter space of our interest to evaluate their exclusion/discovery at the high luminosity LHC.

\section*{Acknowledgements}
A.A. acknowledges partial financial support from the Polish National Science Center under the Beethoven series grant number 2016/23/G/ST2/04301. In the final stage of this project, the research of A.A. has received funding from the Norwegian Financial Mechanism for years 2014-2021, grant Nr DEC-2019/34/H/ST2/00707. R.K.B. thanks the U.S. Department of Energy for the financial support, under grant number DE-SC0016013. Some computing for this project was performed at the High Performance Computing Center at Oklahoma State University, supported in part by the National Science Foundation grant OAC-1531128. R.K.B. thanks Rhitaja Sengupta for helpful discussions. S.~K. is supported by Austrian Science Fund Elise Richter Fellowship  V592-N27 and research group funding FG1. We acknowledge Austrian-India WTZ-DAE exchange project number IN 15/2018. A.D. would like to acknowledge the support of INSPIRE Fellowship IF160414.

\begingroup
\bibliographystyle{unsrturl}
\bibliography{references}
\endgroup

\end{document}